\documentclass[english,12pt,aps,prd,a4paper,preprintnumbers,floatfix,nofootinbib,showpacs,superscriptaddress, notitlepage]{revtex4-1} 
 \pdfoutput=1
\usepackage[usenames,dvipsnames]{color}  
\usepackage{graphicx}
\usepackage{setspace}
\usepackage{braket}
\usepackage{caption}
\usepackage{subcaption}
\usepackage{amsmath}
\usepackage{amssymb}
\usepackage[colorlinks=true,citecolor=darkred,urlcolor=darkred, pdfborder={0 0 0}]{hyperref}
\usepackage[normalem]{ulem}
\usepackage{xcolor}
\usepackage[toc,page]{appendix}
\usepackage[utf8]{inputenc}
\makeatletter
\def\p@subsection{}
\makeatother
\definecolor{grn}{cmyk}{0.53, 0, 0.87, 0.10}
\hypersetup{pdfauthor={Name}}

\usepackage{titlesec}
\titlespacing\section{0pt}{20 pt}{20pt}

\definecolor{darkred}{rgb}{0.6,0,0}

\titleformat*{\subsubsection}{\normalsize \bfseries}
\titleformat*{\subsection}{\normalsize \bfseries}
\titleformat*{\section}{\large \bfseries}

\definecolor{linkcolor}{rgb}{0,0,0.5}

\def\gsim{\raise0.3ex\hbox{$\;>$\kern-0.75em\raise-1.1ex\hbox{$\sim\;$}}}
\def\lsim{\raise0.3ex\hbox{$\;<$\kern-0.75em\raise-1.1ex\hbox{$\sim\;$}}}

\def\beqn#1{\begin{equation}\label{#1}}
\def\eeqn{\end{equation}}

\def\beqa#1{\begin{eqnarray}\label{#1}}
\def\eeqa{\end{eqnarray}}

\def\Z2{$\mathcal{Z_2}$}

\newcommand {\ignore}[1]{}

\def\321{$\mathrm{SU(3) \otimes SU(2) \otimes U(1)}$ }

\newcommand{\AddrIMSC}{
  The Institute of Mathematical Sciences, Chennai 600 113, India}
\newcommand{\AddrCMI}  {Chennai Mathematical Institute, Siruseri 603 103, India}

\newcommand{\AddrIISERB}{Department of Physics, Indian Institute of Science Education and Research - Bhopal, \\ 
Bhopal Bypass Road, Bhauri, Bhopal 462066, India}

\newcommand{\AddrIISERP}{Department of Physics, Indian Institute of Science Education and Research - Pune, \\
Pune, India }

\begin{document}
  
\title{\color{BrickRed} Can Leptonic Mixing Matrix  have a Wolfenstein Form? }
\author{Ankur Panchal}\email{panchal21@iiserb.ac.in}
\affiliation{\AddrIISERB}
\affiliation{\AddrIISERP}
\author{G. Rajasekaran}\email{graj@imsc.res.in}
\affiliation{\AddrIMSC}
\affiliation{\AddrCMI}
\author{Rahul Srivastava}\email{rahul@iiserb.ac.in}
\affiliation{\AddrIISERB}
 \begin{abstract}
  \vspace{1cm} 
  
 We analyze the possibility of the leptonic mixing matrix having a Wolfenstein form at the Grand Unified Theory scale. The renormalization group evolution of masses and mixing angles from the high scale to electroweak scale, in certain new physics scenarios, can significantly alter the form of the leptonic mixing matrix. In the past it was shown that such significant enhancement implies that the leptonic mixing matrix at high scale can be the same or similar in structure to the quark one. We thoroughly analyze this hypothesis in the light of the latest neutrino oscillation data as well as other constraints such as those coming from neutrinoless double beta decay. We show that such an ansatz, at least within the context of minimal supersymmetric models, is no longer compatible with the latest experimental data.

\end{abstract}
\maketitle
%

\section{Introduction}

The Standard Model(SM) of particle physics has been an incredibly successful theory. Discovery of the 125-GeV scalar, if it is confirmed to be the SM Higgs boson will complete the SM~\cite{ATLAS:2012yve,CMS:2012qbp}. However, in spite of its astounding success we now know that SM cannot be the complete theory of nature. The discovery of neutrino oscillations was one of the conclusive proofs for the shortcoming of the SM~\cite{Super-Kamiokande:1998kpq,SNO:2002tuh}. Ever since the discovery of neutrino oscillations our understanding of the neutrino oscillation parameters and hence in turn that of the leptonic mixing matrix is improving. The precision in measurement of certain mixing parameters has dramatically improved over the last decade~\cite{DayaBay:2012fng,RENO:2012mkc,DayaBay:2018yms}. This implies that neutrino physics is now entering the era of precision physics where the experimental data, in particular from  neutrino oscillation experiments, can be used to rule out new physics models in a much more powerful way.  In this work we confront one of the popular theoretical proposals, namely the possibility of leptonic mixing matrix having a Wolfenstein form at some high energy scale.

In its original form the ansatz hypothesized a  ``High Scale Mixing Unification'' (HSMU) between the lepton and quark mixing matrices~\cite{Mohapatra:2003tw}. Here the unification of the two mixing matrices was hypothesized to happen at some high scale typically chosen to be the scale of Grand Unified Theories (GUTs). The Renormalization - Group (RG) evolution of the leptonic mixing angles and masses can then lead to the values of neutrino oscillation parameters with their experimental 3-$\sigma$ range.
One of the key prediction of the  HSMU hypothesis was prediction of a small yet non-zero value of $\theta_{13}$ leptonic mixing angle \cite{Mohapatra:2003tw,Mohapatra:2005gs,Mohapatra:2005pw,Agarwalla:2006dj,Abbas:2013uqh,Abbas:2014ala,Srivastava:2015tza,Srivastava:2016fhg}. This was due to the fact that $\theta_{13} = 0$ is not a fixed point of the RG flow. Rather the RG evolution of the mixing angles from the high scale naturally leads to a small $\theta_{13}$, a fact later observed in experimental measurements~\cite{DayaBay:2012fng,RENO:2012mkc,DayaBay:2018yms}. After the experimental measurement of $\theta_{13}$ angle, the HSMU hypothesis was revisited for both Dirac and Majorana neutrinos and it was shown that indeed HSMU can be a good candidate proposal for understanding of the neutrino mixing and oscillation phenomenon consistent with the experimental data of that time \cite{Abbas:2013uqh,Abbas:2014ala}. The scale of high energy unification as well as dependence on other parameters was also analyzed in these later works showing that the scale of unification does not need to be necessarily the GUT scale. Still later works expanded the idea further looking at the possibility whether the leptonic mixing matrix has a ``Wolfenstein form'' with hierarchical values of mixing angles at high scale, irrespective of whether or not they are exactly same as quark mixing angles \cite{Abbas:2015vba,Abbas:2016qbl,AbdusSalam:2019hov,Rajasekaran:2019uvd}.    

Overall, HSMU and its Wolfenstein form generalization have been shown to be consistent with successive sets of experimental data  for increasingly precise determination of neutrino oscillation parameters over the past decade. However, as  neutrino physics in entering the era of precision measurements, in this work we revisit 
it again to see if it still remains a viable possibility. 
The work-flow is presented in the rest of the paper in the following manner. Section \ref{sec_hsmu} discusses the general framework of HSMU and RG evolution of neutrino oscillation parameters. Then in sections \ref{diraccase} and \ref{majocase} we show our results of HSMU for Dirac and Majorana neutrinos. We find that the current oscillation data combined with other experimental constraints imply that HSMU ansatz is in severe tension with experiments. Further in section \ref{TC} we test whether or not the threshold corrections improve the negative results for HSMU ansatz. The next section \ref{wolf} expands this unification hypothesis into Wolfenstein ansatz by introducing new free parameters. And finally we see conclusions from all the results' interpretations, section \ref{concl}.

\section{High Scale Mixing Unification Hypothesis}
\label{sec_hsmu}

We start with a general discussion of the HSMU hypothesis and the essential ingredients needed to have a large change in values of leptonic mixing angles over the course of RG evolution from high to low energy scale. 

\subsection{General Framework of HSMU}
\label{HSMU:General Framework}

\begin{figure}[ht]
    \centering
    \begin{subfigure}{.5 \textwidth}
      \centering
      \captionsetup{justification=centering}\includegraphics[width=0.9\linewidth]{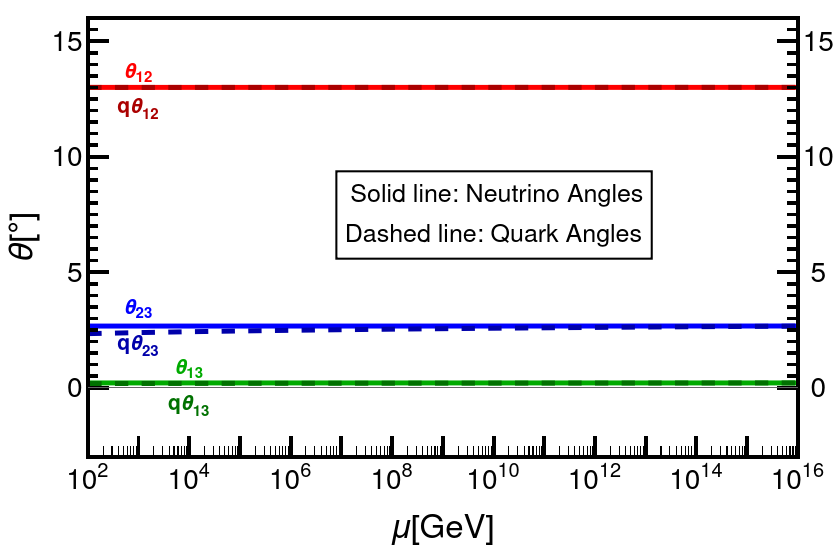}
      \caption{Neutrino and quark mixing angles in SM}
      \label{neutrthetasSM}
    \end{subfigure}%
    \begin{subfigure}{.5\textwidth}
        \centering
        \captionsetup{justification=centering}
      \includegraphics[width=0.9\linewidth]{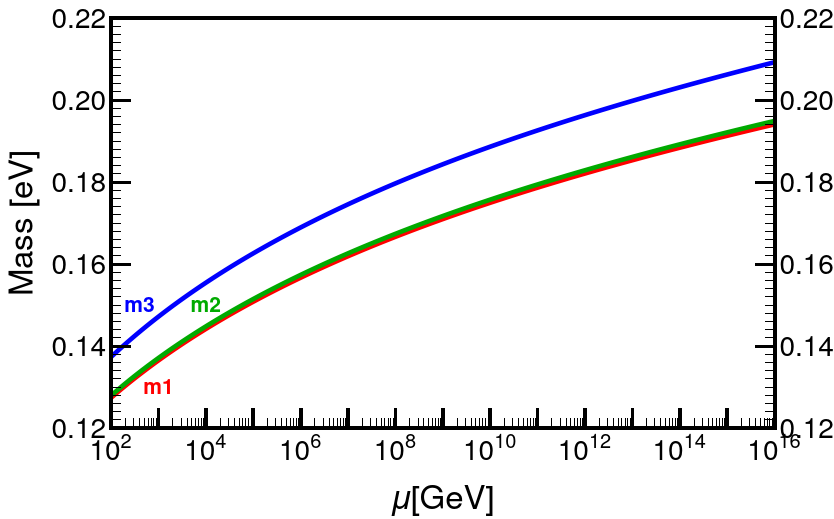}
        \caption{Neutrino masses in SM}
        \label{neutrmassesSM}
    \end{subfigure}
    \caption[RG running of angles]{RG running of neutrino mixing angles and masses in SM from HSMU scale (taken as GUT scale) to low energy scale.}
    \label{rgrunSM}
\end{figure}

Since HSMU assumes that at a high scale, usually taken as GUT scale, the quark and leptonic mixing matrices are one and same, this immediately implies that in order for HSMU to be consistent with neutrino oscillation data, a large change in neutrino mixing angles is needed. Unfortunately this cannot be achieved within SM extended by effective Weinberg operator\footnote{RG evolution including Weinberg operator as effective operator can be done only if its UV cutoff scale is equal to or higher than the HSMU scale.} or it's simple Ultra-Violet (UV) completions such as Type-I seesaw. In Fig.~\ref{rgrunSM} we show the RG evolution of the  mixing angles of quarks and leptons (left panel) and Majorana neutrino masses (right panel) from the HSMU scale (take as GUT scale) to low scale within Type-I seesaw.
As clear from Fig.~\ref{rgrunSM}, within SM + Type-I seesaw the neutrino mixing angles enhance by a negligible amount and hence HSMU hypothesis is completely inconsistent with neutrino oscillation data given in
Tab.~\ref{tab:global-fit}.

\begin{figure}[!ht]
    \centering
    \begin{subfigure}{.5 \textwidth}
      \centering
      \captionsetup{justification=centering}\includegraphics[width=0.9\linewidth]{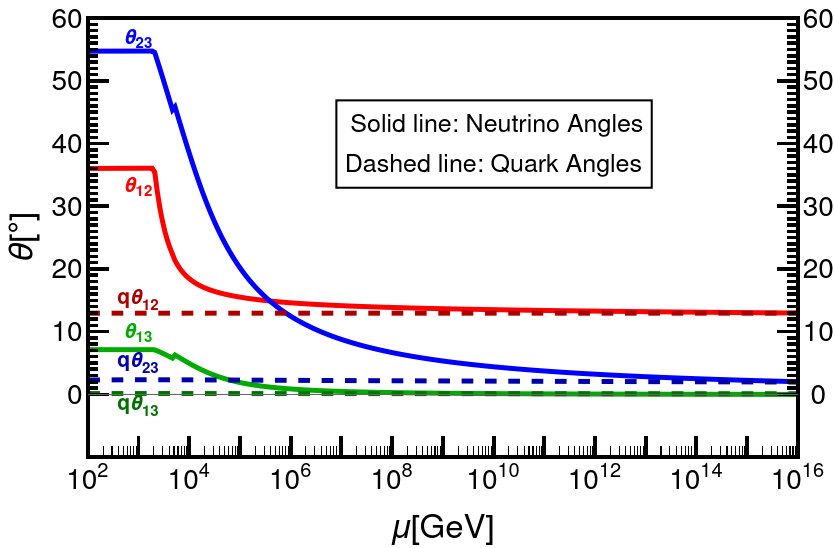}
      \caption{Neutrino and quark mixing angles in MSSM}
      \label{neutrthetasMSSM}
    \end{subfigure}%
    \begin{subfigure}{.5\textwidth}
        \centering
      \captionsetup{justification=centering}
      \includegraphics[width=0.97\linewidth]{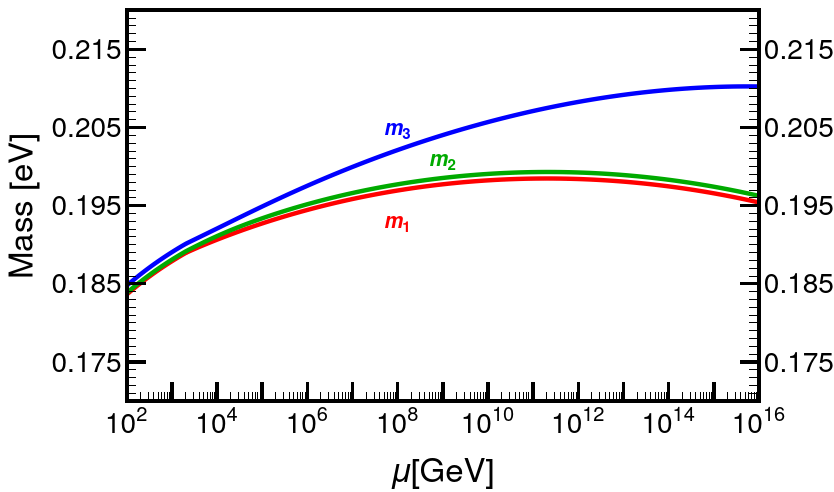}
        \caption{Neutrino masses in MSSM}
        \label{neutrmassesMSSM}
    \end{subfigure}
       \caption[RG running of angles]{RG running of neutrino mixing angles and masses in SM and MSSM rom HSMU scale (taken as GUT scale) to low energy scale. The SUSY breaking scale is taken to be two TeV. }
    \label{rgrunMSSM}
\end{figure}

The situation changes dramatically if there is beyond Standard Model (BSM) new physics at an intermediate scale such as a low scale SuperSymmetry (SUSY) at TeV scale. In Fig.~\ref{rgrunMSSM} we show the RG evolution of neutrino and quark mixing angles (left panel) and the neutrino masses (right panel)  with in Minimal SuperSymmetric Model (MSSM) from HSMU scale to SUSY breaking scale (taken as two TeV) followed by RG evolution in SM from SUSY breaking scale to low scale. 
As can be seen from Fig.~\ref{rgrunMSSM}, in this case a significant enhancement can be achieved during the MSSM part of RG evolution\footnote{Note that since SUSY is typically broken only softly, for RG evolution the details of SUSY breaking are irrelevant. However, in later section we will indeed discuss the leading SUSY threshold corrections and their impact on our analysis.} provided we take large values of $\tan \beta$ and the neutrino masses at HSMU scale are normal ordered (N.O.) and quasi-degenerate~\cite{Mohapatra:2003tw,Mohapatra:2005gs,Mohapatra:2005pw,Agarwalla:2006dj,Abbas:2013uqh,Abbas:2014ala,Srivastava:2015tza,Srivastava:2016fhg}.  In this work we will follow the previous works and  throughout the work we will use MSSM as the intermediate theory with the SUSY breaking scale being two TeV. Moreover, we will always take $\tan \beta = 55$ and the unification scale will always be taken as the GUT scale ($2 \times 10^{16}$ GeV).  The dependence and effect on our analysis due to change in HSMU scale, SUSY breaking scale and $\tan \beta$ were analyzed in detail for Majorana neutrinos in Ref.~\cite{Abbas:2014ala} and for Dirac neutrinos in Ref.~\cite{Abbas:2013uqh}. We will not repeat this analysis here as changing these scales will only increase the tension between the HSMU predictions and current experimental data.

Finally, before going to the sub-cases we must point out that the RG evolution of mixing angles and mass-squared differences are correlated with each other. This is because at high scale we only have three free parameters namely the ``high scale masses'' of the neutrinos using which we have to obtain five neutrino oscillation parameters within their 3-$\sigma$ range at low scale. Furthermore, the RG equations governing the evolution of the mixing parameters are coupled partial differential equations and hence RG evolution of one parameter can strongly effect the RG evolution of the other parameters. As we will see, this means that the values of these parameters at low scale show strong correlations with each other and obtaining all the mixing parameters at low scale within their 3-$\sigma$ ranges is non-trivial.

\subsection{Computational Implementation}

Computationally, the implementation of HSMU can be looked at as a two-step process. First stage involves RG evolution of known values of CKM parameters (see Tab.~\ref{ParticleDataGroup:2018ovx}) from the low scale which we take as the mass scale of Z boson ($\mathrm{M_{Z}}$), to the unification scale which is taken to be the GUT scale.
\begin{table}[!ht]
\begin{center}
    \begin{tabular}{ | c | c | c | }
        \hline
        \hspace{0.1cm} Oscillation parameters\hspace{0.1cm} & 1-$\sigma$ range & \hspace{0.1cm} Best fit values \hspace{0.1cm} \\
        \hline
       $\theta_{q12}$ & \hspace{0.1cm} $12.96^{\circ}$ - $13.04^{\circ}$ \hspace{0.1cm} & $13.00^{\circ}$ \\
        \hline
       $\theta_{q13}$ & $0.20^{\circ}$ - $0.22^{\circ}$ & $0.21^{\circ}$ \\
        \hline
       $\theta_{q23}$ & $2.35^{\circ}$ - $2.45^{\circ}$ & $2.40^{\circ}$ \\
        \hline
        $\delta_q$ & $63.99^{\circ}$-$67.09^{\circ}$ & $65.55^{\circ}$ \\
        \hline
    \end{tabular}
\end{center}
\caption[Low scale Quark oscillation parameters data]{Low scale Quark oscillation parameters data \cite{ParticleDataGroup:2018ovx} $\delta_q$ is CP violation phase of quarks.}
\label{ParticleDataGroup:2018ovx}
\end{table}
While evolving from $\mathrm{M_{Z}}$ scale, we need to provide all known values of Gauge couplings, Yukawa couplings, CP violation phase ($\mathrm{\delta}_{q}$) and low scale quark mixing angles at low scale listed in Appendix~\ref{rgrunning}. Below SUSY breaking scale SM RG equations govern the evolution and above SUSY breaking scale the same is done by MSSM RG equations~\cite{Casas:1999tg,Antusch:2001ck,Antusch:2001vn,Antusch:2002rr,Antusch:2003kp,Antusch:2005gp,Lindner:2005as}.

Now, according to HSMU hypothesis the high scale quark mixing angles are equated with high scale neutrino mixing angles\footnote{We will analyze both cases where the CP phase is also equated and the case where it is taken zero. Majorana phases when taken non-zero are treated as free parameters.}. To evaluate all the neutrino parameters at $\mathrm{M_{Z}}$ scale, we will also need neutrino masses at GUT scale. These are the free parameters of HSMU. With neutrino mixing angles equal to be those of quarks and masses taken as free parameters at the GUT scale, a top-down RG evolution is performed to obtain neutrino oscillation parameters at $\mathrm{M_{Z}}$ scale. We then check the compatibility of the so obtained neutrino oscillation parameters at low scale with the current global fit data listed in Tab.~\ref{tab:global-fit}.
\begin{table}[!ht]
\begin{center}
    \begin{tabular}{ | c |   c | c | }
        \hline
        \hspace{0.1cm} Oscillation parameters \hspace{0.1cm} &  3-$\sigma$ range & \hspace{0.1cm} Best fit values \hspace{0.1cm} \\
        \hline
        $\mathrm{\theta_{12}}$ & \hspace{0.1cm} $31.37^{\circ}$ - $37.41^{\circ}$ \hspace{0.1cm} & $34.33^{\circ}$ \\
        \hline
       $\mathrm{\theta_{13}}$ & $8.13^{\circ}$ - $8.92^{\circ}$ & $8.53^{\circ}$ \\
        \hline
       $\mathrm{\theta_{23}}$ & $41.21^{\circ}$ - $51.35^{\circ}$ & $49.26^{\circ}$ \\
        \hline
        $\Delta \mathrm{m}^2_{\mathrm{atm}}$ ($10^{-3}\mathrm{eV}^2$)&  $2.47$-$2.63$ & $2.55$ \\
        \hline
        $\Delta \mathrm{m}^2_{\mathrm{sol}}$ ($10^{-5}\mathrm{eV}^2$) & $6.94$-$8.14$ & $7.50$\\
        \hline
    \end{tabular}
\end{center}
\caption[Low scale neutrino oscillation parameters data (N.O.)]{Global fit ranges for the neutrino oscillation parameters taking normal mass ordering (N.O.) of the neutrinos \cite{deSalas:2020pgw}. }
\label{tab:global-fit}
\end{table}

 In case of Majorana mass generation, an effective dimension-5 operator is added in the Lagrangian below seesaw scale. Above the seesaw scale, it's UV-completed using type-I seesaw mechanism. All right handed neutrinos added are integrated out below the seesaw scale. During implementation of the RG equations, we have ensured to use the  
appropriate RG equations above and below the SUSY cutoff scale as well as to correctly integrate out the right handed neutrinos for RG evolution below their mass threshold.
All the RG runnings in this work are performed with the help of the Mathematica based package REAP~\cite{Antusch:2005gp}.


\section{Dirac case}
\label{diraccase}


We begin our detailed analysis starting with the case of Dirac neutrinos. If neutrinos are Dirac fermions then by definition, one must add three right handed neutrinos, one for each generation, to the SM particle content. The simplest model for mass generation for Dirac neutrinos is through Higgs mechanism where the smallness of neutrino mass is due to small Yukawa couplings\footnote{In literature there exist various other mass generation mechanisms for Dirac neutrinos, interested readers can see Refs.~\cite{Ma:2014qra,Ma:2015mjd,Ma:2015raa,CentellesChulia:2016rms,CentellesChulia:2017koy,CentellesChulia:2018gwr,CentellesChulia:2018bkz,Bonilla:2018ynb,CentellesChulia:2019xky,CentellesChulia:2020dfh} for some of the recent works and Ref.~\cite{Chulia:2021jgv} for a review.}. As mentioned before, for HSMU one needs SUSY at the TeV scale which its simplest form can be implemented by embedding SM in MSSM with three additional superfields embedding the right handed neutrinos. The Lagrangians before and after SUSY breaking scale are given by
\begin{eqnarray}
\text{Below SUSY breaking scale:} && \nonumber\\
 \mathcal{L} & = & \mathcal{L}_{\mathrm{SM}} + \mathcal{L}_{\mathrm{\nu_{R}}} \, \, = \, \mathcal{L}_{\mathrm{SM}} - \mathrm{Y^{ij}_{\nu} \bar{L}^{i} \tilde{H} \nu^{j}_{R}} + h.c. \\
\text{Above SUSY breaking scale:} && \nonumber\\
 \mathcal{L}  & = &  \mathcal{L}_{\mathrm{MSSM}} + \mathcal{L}_{\mathrm{\nu_{R}}} \, \, = \, \mathcal{L}_{\mathrm{MSSM}} - \mathbb{Y}^{ij}_{\nu} \bar{\mathbb{L}}^{i} \mathbb{H}_u \mathbb{N}^{j}_{R} + h.c.
 \label{eq:dir-lags}
\end{eqnarray}
where $i, j, k$ are flavor indices, $\mathrm{Y_{\nu}}$ is the Yukawa matrix for the neutrinos, 
$\mathrm{\nu^{i}_{R}}$ is a right handed neutrino of flavor $i$ and $\mathbb{L}, \mathbb{H}_u, \mathbb{N}$ are the corresponding superfields. The RG equations for the evolution of neutrino masses and mixing parameters for this case can be found in \cite{Lindner:2005as}.

The implementation of HSMU ansatz in this case follows the general strategy discussed in previous section. We start with the know values of the quark masses, mixing parameters, gauge couplings etc at the low scale ($M_Z$). We use the SM RG equations for the evolution up to SUSY breaking scale (2 TeV) and then use the MSSM RG equations up to the high scale (GUT scale). At the GUT scale we fix the leptonic mixing angles and depending on the case (see discussion below) the CP phase also to be equal to the quark ones. The neutrino masses at GUT scale are taken as N.O. and are treated as free parameters. We then do RG running back to the low scale to obtain the RG evolved values of the neutrino oscillation parameters and compare with the global fit data~\cite{deSalas:2020pgw}. 

In this case, we consider two sub-cases: one when there is no CP violation in the leptonic sector i.e the CP phase $\delta = 0$) and the other when there is CP violation in the leptonic sector as well i.e. $\delta \neq 0$. In the second case following HSMU ansatz, at high scale we take the  $\delta = \delta_q$. We analyze both cases one by one. 
%

\subsection{\textbf{CP conserving  $\delta = 0$ case }}

The experimental situation regarding CP violation in leptonic sector is still not clear. Thus there is possibility that there is no CP violation in leptonic sector, this implies that the CP phase $\delta = 0$. The CP phase being zero is a fixed point in RG evolution which means that if $\delta = 0$ at high scale it will remain zero at the low scale as well. Thus, for the case of CP conservation in leptonic sector, one must take $\delta = 0$ at the HSMU scale. 

\begin{figure}[!ht]
    \centering
    \includegraphics[width=0.7\linewidth]{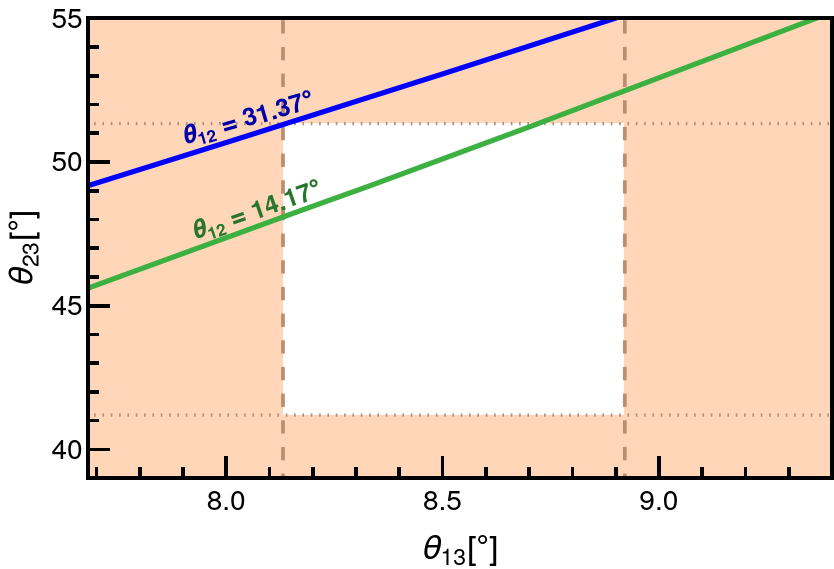}
    \caption[Dirac case($\delta = 0^{\circ}$): $\theta_{23}$ vs $\theta_{13}$ correlation]{$\theta_{23}$ vs $\theta_{13}$ with $\delta = 0^{\circ}$. The correlation is a straight line. Different straight lines would correspond to different $\theta_{12}$. The shaded region represents rejected values of $\theta_{23}$ and $\theta_{13}$ which are out of their 3-$\sigma$ ranges. }
    \label{diraccp0}
\end{figure}

As mentioned before, within HSMU hypothesis the RG evolution of neutrino oscillation parameters are correlated. In Fig.~\ref{diraccp0} we show the obtained values of $\theta_{23}$ and $\theta_{13}$ at low scale for benchmark values of $\theta_{12}$ angle. 
The shaded region(light red) corresponds to values of $\theta_{23}$ and $\theta_{13}$ outside their 3-$\sigma$ range given in Tab.~\ref{tab:global-fit}. The white rectangular region corresponds to allowed values of both $\theta_{23}$ and $\theta_{13}$ simultaneously. For the correlation line passing through the allowed region, $\theta_{12}$ is $14.17^\circ$ which is an outside 3-$\sigma$ low scale value for $\theta_{12}$. Thus this line can be discarded as we can not have the low scale values of all three angles within their 3-$\sigma$ ranges. When $\theta_{12}$ is varied the linear correlation function moves with respect to the allowed (white) region. If $\theta_{12}$ is increased the line moves to the top-left corner of the allowed region, indicating that for those values of $\theta_{12}$ we can not have $\theta_{23}$ and $\theta_{13}$ within their 3-$\sigma$ ranges. We can see that for the minimum value of $\theta_{12}$ inside its 3-$\sigma$ range,  the line is completely outside the allowed region for $\theta_{23}$ and $\theta_{13}$, which tells us that we cannot have a suitable case where all of the mixing angles lie within their valid ranges. This happens because RG equations are monotonous functions of $\theta_{12}$ and therefore any further increment in the value of $\theta_{12}$ will only move the correlation further away from the allowed region.

\subsection{\textbf{CP violating $\delta = \delta_q$ case}}

\begin{figure}[!ht]
    \centering
    \includegraphics[width=0.7\linewidth]{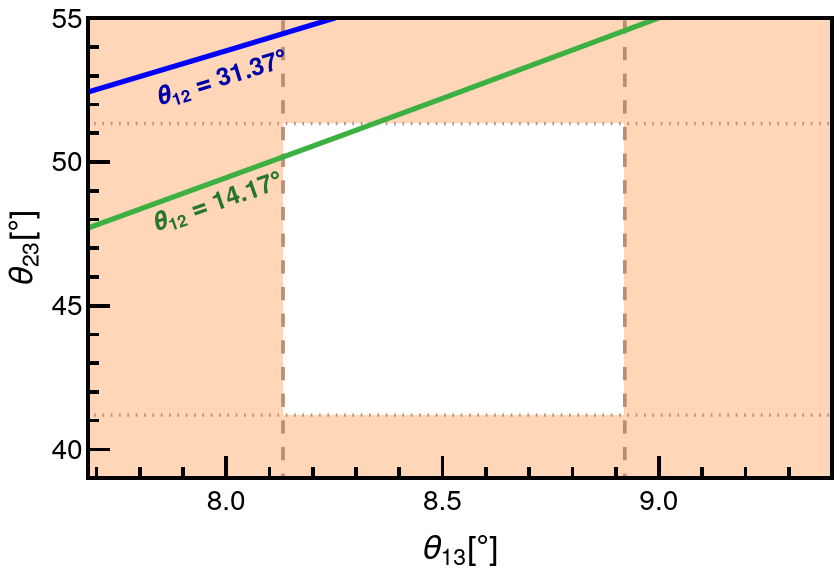}
    \caption[Dirac case($\delta \neq 0^{\circ}$): $\theta_{23}$ vs $\theta_{13}$ correlation]{$\theta_{23}$ vs $\theta_{13}$ with $\delta \neq 0^{\circ}$. At GUT scale the value of $\delta$ is set equal to $\delta_q$ at that scale. The shaded region represents the rejected values as before. }
    \label{diraccpqdel}
\end{figure}

In this case we take $\delta$ at high scale to be equal to quark sector CP violating phase at high scale i.e.  $\delta = \delta_q$ at the HSMU scale. Running down from GUT scale to $\mathrm{M_{Z}}$ scale, we still can't have all three angles inside their 3-$\sigma$ ranges, as can be seen from Fig.~\ref{diraccpqdel}. The result remains the same as in CP conserving case. Valid values of $\theta_{12}$ will correspond to $\theta_{23}$ vs $\theta_{13}$ correlation line lying completely outside the allowed region.

Thus, for the case of Dirac neutrinos neither CP conserving nor CP violating cases allow us to have all mixing angles within their 3-$\sigma$ range. Since mixing angles don't agree with the experimental values for Dirac case, it is redundant to check the same for mass squared differences. hence, we can conclude that although HSMU for Dirac neutrinos was a valid possibility in the past~\cite{Abbas:2013uqh}, the current more stringent data rules it out as a viable possibility.


\section{Majorana case} 
\label{majocase}


In this case we consider neutrinos to be Majorana particles whose mass is generated through Type-I seesaw mechanism. Current limits on neutrino mass~\cite{Planck:2018vyg,KATRIN:2021uub} imply that if all Yukawas are taken within their perturbative range then the Type-I seesaw scale has to be smaller than the GUT scale. Throughout this work we will take the seesaw scale to be $\sim 10^{12}$ GeV so that the neutrino Yukawa couplings remain throughout the entire range of RG running. In this case there are several scales and the Lagrangian used for RG running at each scale is shown in \eqref{eq:maj-lags}.
\begin{eqnarray}
  \text{Below SUSY breaking scale:} & &\nonumber\\
 \mathcal{L} &  = &  \mathcal{L}_{\mathrm{SM}} + \mathcal{L}_{5} \nonumber \\
 \text{SUSY  up to  seesaw scale:} & & \nonumber \\
 \mathcal{L} & = &  \mathcal{L}_{\mathrm{MSSM}} + \mathcal{L}_{5}  \\
 \text{Seesaw to HSMU scale:} & & \nonumber \\
  \mathcal{L} & = &  \mathcal{L}_{\mathrm{MSSM}} + \mathcal{L}_{\mathrm{seesaw}} 
 \, \,  = \, \mathcal{L}_{\mathrm{MSSM}} - \mathbb{Y}^{ij}_{\nu} \bar{\mathbb{L}}^{i} \mathbb{H}_u \mathbb{N}^{i}_{R} - \frac{1}{2} \mathbb{M}^{ij}  (\bar{\mathbb{N}}^c)^{j} \mathbb{N}^{j} + \mathrm{h.c.}  \nonumber
 \label{eq:maj-lags}
\end{eqnarray}
where $\mathbb{M}$ represents the Majorana mass matrix, $\mathbb{Y}$ is the Yukawa matrix and rest of the notation remains same as in \eqref{eq:dir-lags}. Furthermore, in \eqref{eq:maj-lags}, below seesaw scale, we have added the dimension five Weinberg operator ($\mathcal{L}_{5}$) obtained from integrating out the right handed neutrinos~\cite{Weinberg:1979sa}. In its SUSY version it is given by
\begin{equation} \label{l5op}
 \mathcal{L}_{5} = - \frac{\mathbb{\kappa}^{ij}}{\Lambda}(\bar{\mathbb{L}}^c)^{i} \mathbb{H}_d \mathbb{H}_{d} \mathbb{L}^j
\end{equation}
where $\Lambda$ is the cutoff scale for the effective operator which in our case is the seesaw scale and $\mathbb{\kappa}^{ij}$ is the effective coupling. After SUSY breaking the $\mathcal{L}_5$ becomes the canonical Weinberg operator added to SM Lagrangian~\cite{Weinberg:1979sa}.

At the HSMU scale, apart from the masses of neutrinos we now have additional three free parameters, namely the three phases, the usual $\delta$ CP as well as the two  Majorana phases $\mathrm{\varphi}_1$ and $\mathrm{\varphi}_2$. Here again we have to consider two sub-cases- one when Majorana phases are zero and the other when they are non-zero. For simplicity, we are fixing $\delta$ to be equal to $\delta_{q}$ at high scale. We have check and verified that the case of no CP violation in Majorana case also leads to similar qualitative results as the case of 
$\delta \delta_{q}$ and $\mathrm{\varphi}_1 = \mathrm{\varphi}_2 = 0$. Therefore, in order to avoid unnecessary repetition, we will not present it separately. 

Before looking at the sub-cases we should point out that for Majorana neutrinos one also has an additional constraint coming from the experimental searches of neutrinoless double beta decay ($0\nu \beta\beta$). The $0\nu \beta \beta$ experiments provide constraints on a particular combination of neutrino masses and mixing parameters called
the effective Majorana mass ($\mathrm{m_{\beta \beta}}$) which is given as 
\begin{equation}
 \mathrm{m_{\beta \beta}} =  |c^{2}_{12}c^{2}_{13}m_{1} + s^{2}_{12}c^{2}_{13}m_{2} e^{i \varphi_{1}} + s^{2}_{13}m_{3} e^{i \varphi_{2}}|
\end{equation}
where $\mathrm{c_{ij} = \cos{\theta_{ij}}}$ and $\mathrm{s_{ij} = \sin{\theta_{ij}}}$ denote the sine and cosine of the mixing angles.
 
The decay rate of $0 \nu \beta \beta$ process $\Gamma \propto | m_{\beta \beta} |^2$. The most stringent upper bound range on $\mathrm{m_{\beta \beta}}$ is set by KamLAND-Zen experiment and its value ranges from $0.065$ eV to $0.165$ eV \cite{KamLANDZen} depending on the choice of the nuclear matrix elements. In this paper we will use the conservative upper bound, i.e., we will demand that the low scale value of $\mathrm{m_{\beta \beta}} < 0.165$ eV.


\subsection{\textbf{Majorana phases,  $\mathrm{\varphi}_1$ $=$ $\mathrm{\varphi}_2$ $= 0^{\circ}$}}

\begin{figure}[!ht]
    \centering
    \includegraphics[width=0.7\linewidth]{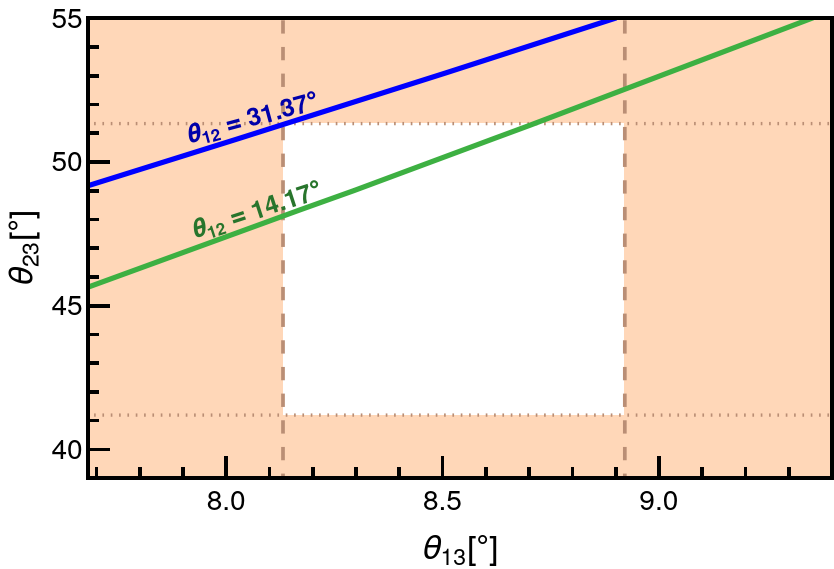}
    \caption[Majorana case($\varphi_{1}=\varphi_{2}=0^{\circ}$): $\theta_{23}$ vs $\theta_{13}$ correlation]{$\theta_{23}$ vs $\theta_{13}$ with $\mathrm{\varphi}_1$ $=$ $\mathrm{\varphi}_2$ $= 0^{\circ}$. The shaded region are outside the 3-$\sigma$ range.}
    \label{fig:majo0}
\end{figure}

This case as well as the case of no leptonic CP violation are both qualitatively similar to Dirac cases of no leptonic CP violation and $\delta = \delta_q$, respectively. Like the Dirac case here again we first look at the correlated evolution of the mixing angles at the low scale. In Fig.~\ref{fig:majo0} we show the correlation between low scale value of $\theta_{23}$ and $\theta_{13}$ mixing angles for benchmark choices of the $\theta_{12}$ angle.
Just like the Dirac case, here also, the three mixing angles cannot be simultaneously brought within their current 3-$\sigma$ range. Since all three angles couldn't be brought into their 3-$\sigma$ ranges, there is no reason to look into mass squared differences or $\mathrm{m_{\beta \beta}}$'s experimental constraints.  To conclude this case is also ruled out by the current experimental data.

\subsection{\textbf{Non-zero Majorana phases, $\varphi_{1} \neq 0, \varphi_{2} \neq 0$}}

\begin{figure}[!ht]
    \centering
    \includegraphics[width=0.7\linewidth]{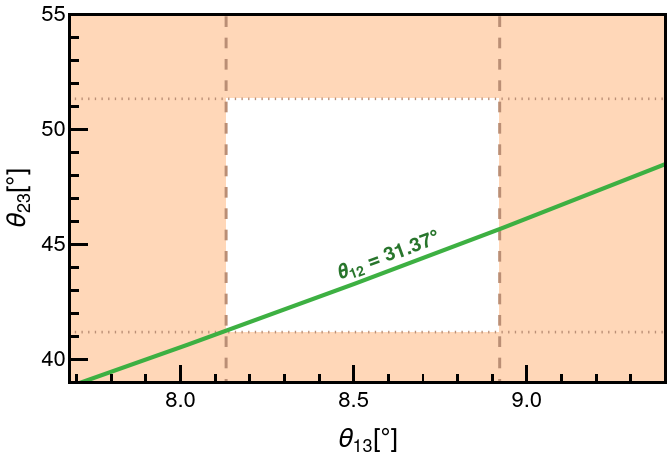}
    \caption[Majorana case($\varphi_{1}=120^{\circ},\varphi_{2}=0^{\circ}$): $\theta_{23}$ vs $\theta_{13}$ correlation]{$\theta_{23}$ vs $\theta_{13}$ with $\mathrm{\varphi}_1$ $= 120^{\circ}$, $\mathrm{\varphi}_2$ $= 30^{\circ}$. Shaded regions are outside the 3-$\sigma$ range. }
    \label{fig:majononzero}
\end{figure}

In this case, apart from the neutrino masses at HSMU scale, we have two more free parameters namely the two Majorana phases, $\varphi_{1}, \varphi_{2}$ which cannot be constrained by the HSMU hypothesis. Non-zero values of the Majorana phases at HSMU scale strongly influence the RG evolution of the neutrino oscillation parameters. In fact, contrary to the previous cases here by appropriate choices of the two Majorana phases, one can indeed simultaneously bring all the three mixing angles inside their 3-$\sigma$ ranges. In Fig.~\ref{fig:majononzero} we show the correlation between $\theta_{23}$ and $\theta_{13}$ for a value of $\theta_{12}$ inside its 3-$\sigma$ range and for benchmark choices of the Majorana phases $\varphi_{1}, \varphi_{2}$.

Therefore this case needs closer inspection to see if there exist values of the Majorana phases for which the mass square differences ($\Delta \mathrm{m}^2_{\mathrm{atm}}$ and $\Delta \mathrm{m}^2_{\mathrm{sol}}$) can also be brought within their 3-$\sigma$ range.
In addition  we also need to check if the experimental upper allowed limit of  $\mathrm{m_{\beta \beta}}$ is also respected.
Thus one should check for all possible combinations of values of $\varphi_{1}, \varphi_{2}$. This is computationally very challenging. Fortunately, we don't have to as the dependence of RG evolution of the mixing parameters on the Majorana phases is not completely arbitrary.  
Thus, instead of scanning through randomly generated values of all possible combinations of $\mathrm{\varphi}_1$, $\mathrm{\varphi}_2$ for the whole [0,2$\pi$] range of both phases, we manually choose some discrete benchmark values of them which clearly show the RG evolution pattern.
To this end we fix one $\mathrm{\varphi}$ and record the effect of other $\mathrm{\varphi}$ on the low scale mixing angles as well as on $\Delta \mathrm{m}^2_{\mathrm{atm}}$ \& $\Delta \mathrm{m}^2_{\mathrm{sol}}$ as we discuss now.

\subsubsection{Effect of $\mathrm{\varphi}_1, \mathrm{\varphi}_2$ on $\Delta \mathrm{m}^2_{\mathrm{atm}}$, $\Delta \mathrm{m}^2_{\mathrm{sol}}$, $\mathrm{m_{\beta \beta}}$ and $\mathrm{m_{lightest}}$ :}

\begin{figure}[!ht]
    \centering
    \vspace{-.7cm}
  \hspace{-55 pt}  \begin{subfigure}{.46 \textwidth}
      \centering
      \captionsetup{oneside,margin={0.8cm,0cm}}
      \includegraphics[width=\linewidth]{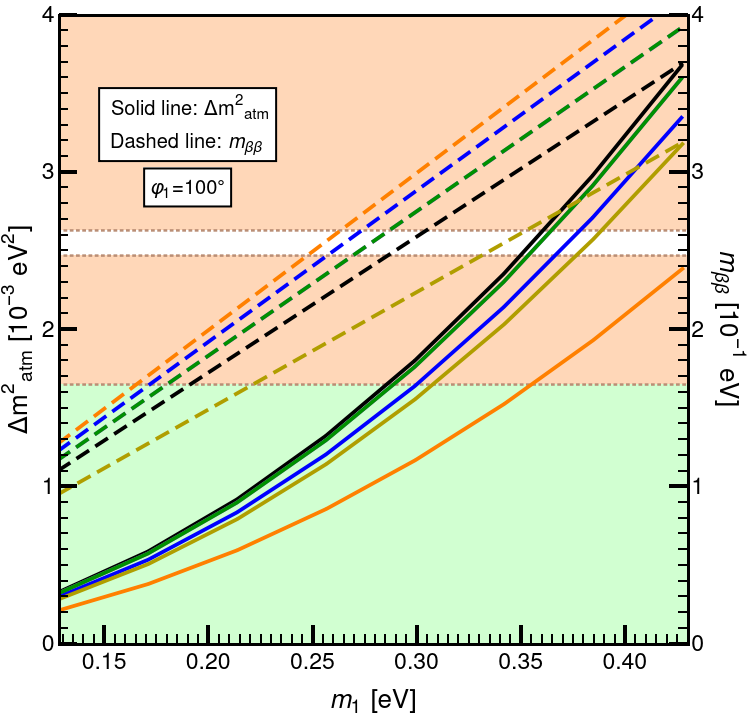}
      \caption{$\Delta \mathrm{m^{2}_{atm}}$ vs $\mathrm{m_{lightest}}$ for $ \varphi_{1} = 100^{\circ}$}
      \label{atmvsmlight_100}
    \end{subfigure}%
    \hspace{20 pt}
    \begin{subfigure}{.5\textwidth}
        \centering
         \captionsetup{oneside,margin={0.8cm,0cm}}
        \includegraphics[width=1.2\linewidth]{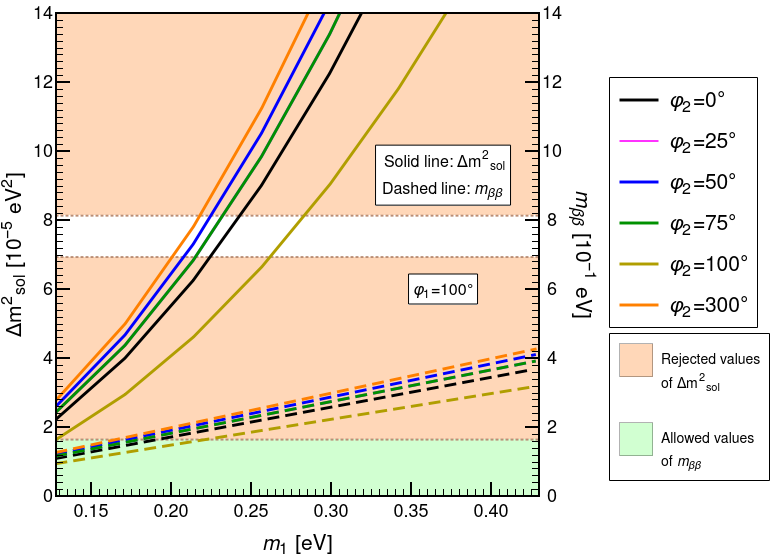}
        \caption{$\Delta \mathrm{m^{2}_{sol}}$ vs $\mathrm{m_{lightest}}$ for $ \varphi_{1} = 100^{\circ}$}
        \label{solvsmlight_100}
    \end{subfigure} \\
     \hspace{-55 pt}
     \begin{subfigure}{.46 \textwidth}
      \centering
      \captionsetup{oneside,margin={0.8cm,0cm}}
      \includegraphics[width=\linewidth]{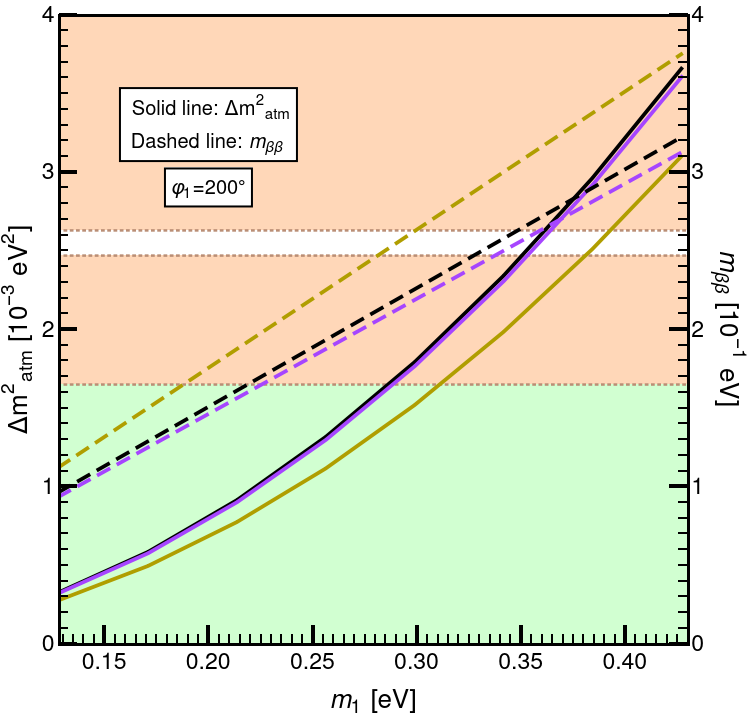}
      \caption{$\Delta \mathrm{m^{2}_{atm}}$ vs $\mathrm{m_{lightest}}$ for $ \varphi_{1} = 200^{\circ}$}
      \label{atmvsmlight_200}
    \end{subfigure}%
    \hspace{20 pt}
    \begin{subfigure}{.5\textwidth}
        \centering
      \captionsetup{oneside,margin={0.8cm,0cm}}        \includegraphics[width=1.2\linewidth]{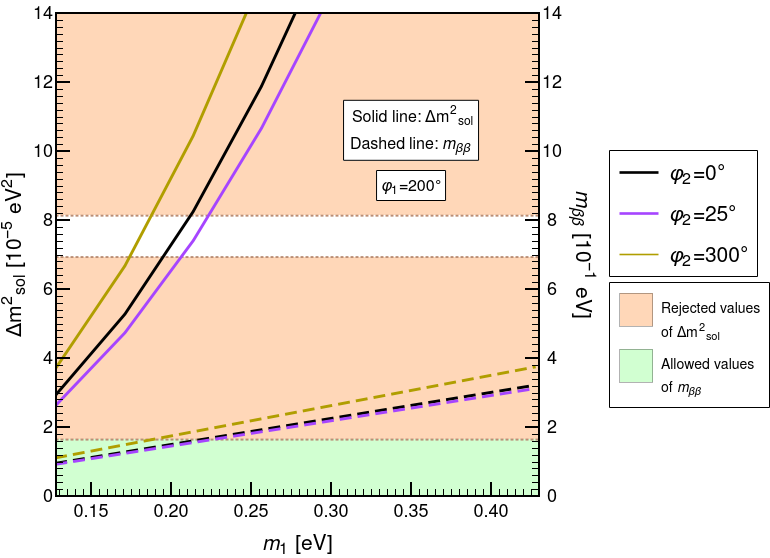}
        \caption{$\Delta \mathrm{m^{2}_{sol}}$ vs $\mathrm{m_{lightest}}$ for $ \varphi_{1} = 200^{\circ}$}
        \label{solvsmlight_200}
    \end{subfigure} \\
     \hspace{-55 pt}
    \begin{subfigure}{.46 \textwidth}
     \centering
      \captionsetup{oneside,margin={0.8cm,0cm}}
      \includegraphics[width=\linewidth]{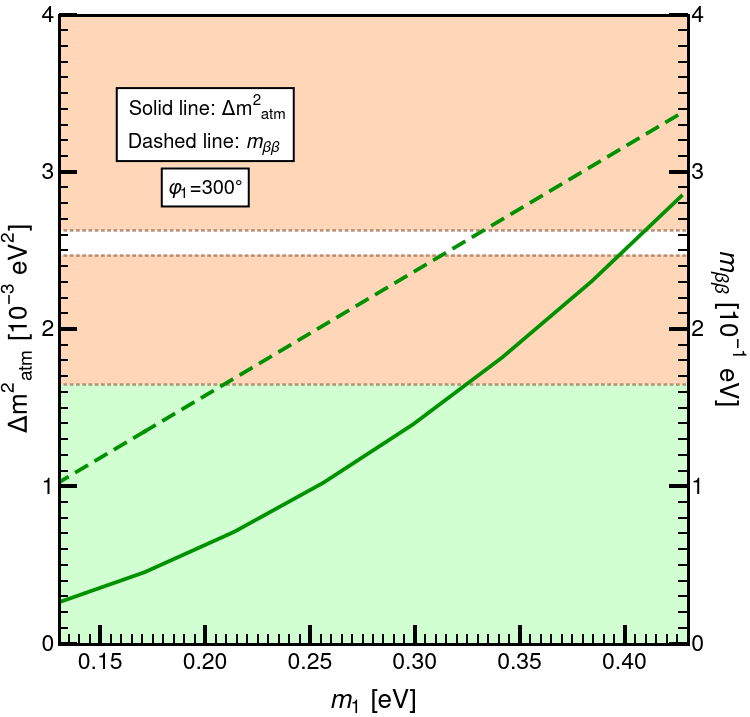}
      \caption{$\Delta \mathrm{m^{2}_{atm}}$ vs $\mathrm{m_{lightest}}$ for $ \varphi_{1} = 300^{\circ}$}
      \label{atmvsmlight_300}
    \end{subfigure}%
   \hspace{20 pt}
    \begin{subfigure}{.5\textwidth}
     \centering
      \captionsetup{oneside,margin={0.8cm,0cm}}       \includegraphics[width=1.2\linewidth]{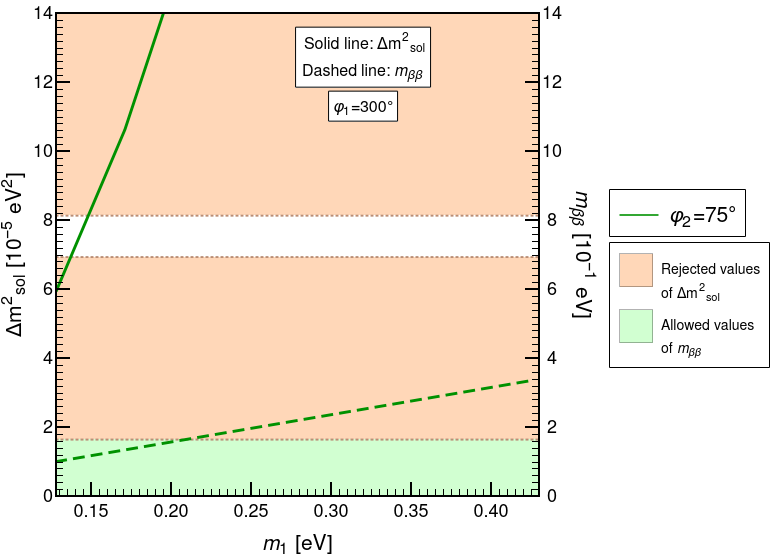}
        \caption{$\Delta \mathrm{m^{2}_{sol}}$ vs $\mathrm{m_{lightest}}$ for $ \varphi_{1} = 300^{\circ}$}
        \label{solvsmlight_300}
    \end{subfigure}
\caption{ \begin{footnotesize} The correlated evolution of $\Delta \mathrm{m}^2_{\mathrm{atm}}$ (left panels) and $\Delta \mathrm{m}^2_{\mathrm{sol}}$ (right panels) with respect to $\mathrm{m_{lightest}}$ for different benchmark values of the Majorana phases $\mathrm{\varphi}_1, \mathrm{\varphi}_2$. Also plotted are the values of $\mathrm{m_{\beta \beta}}$ with respect to $\mathrm{m_{lightest}}$. Solid lines represent $\Delta \mathrm{m}^2_{\mathrm{atm}}$ or $\Delta \mathrm{m}^2_{\mathrm{sol}}$ and dotted lines represent $\mathrm{m_{\beta \beta}}$. The white bands in the plots show the 3-$\sigma$ range of the mass square differences. Furthermore, in all the plots we have kept the three mixing angles always inside their 3-$\sigma$ range.  See text for more details.                                                                                                                                                                                                                                                                                                                                                                                                                                                                                                                                                                                                                                                                                                                                                                                 \end{footnotesize}}
\label{mlight}
\end{figure}

The Majorana phases and neutrino masses are the only free parameters at the HSMU scale and their choices play a critical role in RG evolution of the neutrino oscillation parameters. At low scale, apart from the mixing angles one has to ensure that the mass square differences $\Delta \mathrm{m^{2}_{atm}}$, $\Delta \mathrm{m^{2}_{sol}}$ remain in their 3-$\sigma$ range and $\mathrm{m_{\beta \beta}}$ remains below its upper limit. It can be insightful to understand the extent by which we can bringing $\Delta \mathrm{m^{2}_{atm}}$ , $\Delta \mathrm{m^{2}_{sol}}$ and $\mathrm{m_{\beta \beta}}$ in their valid ranges. This  can be analyzed by plotting them against one of the masses at the low scale. We choose the lightest neutrino mass ($\mathrm{m_{lightest} = m_{1}}$) for the same. Fig.\ref{mlight} shows the trend of $\Delta \mathrm{m^{2}_{atm}}$ , $\Delta \mathrm{m^{2}_{sol}}$ and $\mathrm{m_{\beta \beta}}$ plotted against $\mathrm{m_{lightest}}$ with respect to the variations of $\mathrm{\varphi}_1$ , $\mathrm{\varphi}_2$.
%

In Fig.~\ref{mlight} the solid lines represent $\Delta \mathrm{m}^2_{\mathrm{atm}}$ or $\Delta \mathrm{m}^2_{\mathrm{sol}}$ and dotted lines represent $\mathrm{m_{\beta \beta}}$. The white bands in the plots show the 3-$\sigma$ range of the mass square differences. Furthermore, in all the plots we have kept the three mixing angles always inside their 3-$\sigma$ range.  
In the various panels of Fig.~\ref{mlight} we can notice the region on $\mathrm{m_{lightest}}$ axis where the $\mathrm{m_{\beta \beta}}$ dashed lines are below their upper bound. The solid lines for those $\mathrm{m_{\beta \beta}}$ values can be inside the white region or outside the white region. Comparing graphs for $\Delta \mathrm{m}^2_{\mathrm{sol}}$ and $\Delta \mathrm{m}^2_{\mathrm{atm}}$ side by side one can estimate whether, for a particular set of Majorana phases or for a particular value of $\mathrm{m_{lightest}}$, whether $\Delta \mathrm{m}^2_{\mathrm{sol}}$ or $\Delta \mathrm{m}^2_{\mathrm{atm}}$ are above or below their corresponding valid values.
This can help us select such a data set which has either of $\Delta \mathrm{m}^2_{\mathrm{sol}}$ or $\Delta \mathrm{m}^2_{\mathrm{atm}}$ inside the white region and the other one very near to its range. And then based on which mass sqaure difference is outside by what extent, one can determine how to bring in the fifth low scale parameter value inside its 3-$\sigma$ range.

Consider Fig.\ref{atmvsmlight_100} and Fig.\ref{solvsmlight_100}. In both the figures, for no values of $\mathrm{m_{lightest}}$($\mathrm{m_1}$), $\Delta \mathrm{m^2}$s and $\mathrm{m_{\beta \beta}}$ lie in their allowed ranges simultaneously. In conclusion, for this set of $\mathrm{\varphi}$'s, we can not have both $\Delta \mathrm{m}^2_{\mathrm{sol}}$ and $\Delta \mathrm{m}^2_{\mathrm{atm}}$ inside their 3-$\sigma$ ranges satisfying the $\mathrm{m_{\beta \beta}}$ constraint and thus we can discard this set of $\mathrm{\varphi}_1, \mathrm{\varphi}_1$ values.
From Fig.\ref{atmvsmlight_200}, we can see that there are only 3 combinations of $\mathrm{\varphi}_1$ and $\mathrm{\varphi}_2$ for which all three neutrino angles could be brought in. But again for valid $\mathrm{m_{\beta \beta}}$ values, there is no value of $\mathrm{m_{lightest}}$ where both $\Delta \mathrm{m}^2_{\mathrm{sol}}$ and $\Delta \mathrm{m}^2_{\mathrm{atm}}$ can simultaneously lie within their 3-$\sigma$ ranges. Thus this set of values can also be rejected
The discussion for  Fig.\ref{atmvsmlight_300} and  Fig.\ref{solvsmlight_300} is pretty much the same, except for the fact that there is only one pair of $\mathrm{\varphi}_1$ and $\mathrm{\varphi}_2$, out of the chosen ones, brings all angles inside. But again $\Delta \mathrm{m}^2_{\mathrm{sol}}$, $\Delta \mathrm{m}^2_{\mathrm{atm}}$ and 
$\mathrm{m_{\beta \beta}}$ cannot all be brought simultaneously within their allowed ranges for any choice of $\mathrm{m_{lightest}}$.

Similarly, we have searched for many possible combinations of Majorana phases some of which are summarized in Tab.~\ref{checktable} which shows whether or not the low scale parameters are  within their 3-$\sigma$ ranges for each pair of $\mathrm{\varphi}_1$, $\mathrm{\varphi}_2$ examined. From Table \ref{checktable}, it is clear that with $\mathrm{m_{\beta \beta}}$ constraint applied, we can not bring in $\Delta \mathrm{m}^2_{\mathrm{atm}}$ and $\Delta \mathrm{m}^2_{\mathrm{sol}}$, along with all 3 angles, inside their allowed ranges simultaneously for any combination of high scale $\mathrm{\varphi}$'s. It is to be noted that in spite of examining only a few discrete values of $\mathrm{\varphi}$'s, we can claim this for the whole range [0,2$\pi$] because $\Delta \mathrm{m}^2_{\mathrm{atm}}$ varies monotonously and continuously with $\mathrm{\varphi}$s. 

\begin{table}[!ht]
    \centering
        \begin{tabular}{ |p{30 pt}|p{30 pt}|p{20 pt}|p{20 pt}|p{20 pt}|p{35 pt}|p{40 pt}| }
 \hline
$\varphi_{1}(^{\circ})$ & $\varphi_{2}(^{\circ})$  &  $\theta_{12}$ & $\theta_{13} $ & $\theta_{23} $ & $\Delta m^{2}_{sol} $ & $\Delta m^{2}_{atm} $\\ \hline \hline
 50 & 0 & \checkmark & \checkmark  & \checkmark  & \checkmark  & $\times$  \\ \hline
 100 & 0 & \checkmark & \checkmark  & \checkmark  & $\times$  & $\times$  \\ \hline
 200 & 0 & \checkmark & \checkmark & \checkmark & $\times$ & $\times$ \\ \hline
 300 & 0 & \checkmark & $\times$ & \checkmark & \checkmark & $\times$  \\ \hline
 0 & 50 & \checkmark & $\times$ & \checkmark & \checkmark & $\times$ \\ \hline
 50 & 50 & \checkmark & $\times$ & \checkmark & \checkmark & $\times$  \\ \hline
 100 & 50 & \checkmark & \checkmark  & \checkmark  & $\times$  & $\times$  \\ \hline
 200 & 50 & \checkmark & $\times$  & \checkmark  & $\times$  & $\times$  \\ \hline
 300 & 50 & \checkmark & \checkmark  & \checkmark  & \checkmark  & $\times$  \\ \hline
 0 & 100 & \checkmark & $\times$ & \checkmark & \checkmark & $\times$  \\ \hline
 50 & 100 & \checkmark & $\times$ & \checkmark & \checkmark & $\times$  \\ \hline
 100 & 100 & \checkmark & \checkmark  & \checkmark  & $\times$  & $\times$  \\ \hline
 200 & 100 & \checkmark & $\times$  & \checkmark  & $\times$  & $\times$  \\ \hline
 300 & 100 & \checkmark & $\times$ & \checkmark & \checkmark & $\times$  \\ \hline
 0 & 200 & \checkmark & $\times$  & $\times$  & $\times$  & $\times$  \\ \hline
 50 & 200 & \checkmark & $\times$ & \checkmark & \checkmark & $\times$  \\ \hline
 100 & 200 & \checkmark & $\times$  & \checkmark  & $\times$  & $\times$  \\ \hline
 200 & 200 & \checkmark & $\times$  & \checkmark  & $\times$  & $\times$  \\ \hline
 300 & 200 & \checkmark & \checkmark  & $\times$  & $\times$  & \checkmark  \\ \hline
 0 & 300 & \checkmark & $\times$ & \checkmark & \checkmark & $\times$  \\ \hline
 50 & 300 & \checkmark & \checkmark  & \checkmark  & \checkmark  & $\times$  \\ \hline
 100 & 300 & \checkmark & \checkmark  & \checkmark  & $\times$  & $\times$  \\ \hline
 200 & 300 & \checkmark & \checkmark  & \checkmark  & \checkmark  & $\times$  \\ \hline
 300 & 300 & \checkmark & $\times$ & \checkmark & \checkmark & $\times$  \\ \hline

\end{tabular} 

    \caption[Low scale neutrino parameters vs $\varphi_{1}, \varphi_{2}$]{` \checkmark ' represents that the corresponding parameter is within its experimental 3-$\sigma$ range. $\mathrm{m_{\beta \beta}}$ is inside its bounds for ALL combinations of $\mathrm{\varphi}_1$, $\mathrm{\varphi}_2$. In some cases, we can bring one of the mass square difference in at the expense of $\mathrm{m_{\beta \beta}}$ being outside its bound.
    }
    \label{checktable}
\end{table}

Thus, we can conclude that although this case initially appeared to be promising but on closer inspection we found that this case also doesn't lead to any viable parameter space where HSMU ansatz in compatible with the current experimental data. However, before completely rejecting the HSMU ansatz we need to do one final check namely the effect of SUSY threshold corrections on the low scale values of neutrino oscillation parameters, which we do in the next section.

\section{Low Energy SUSY Threshold Corrections}
\label{TC}

As we can see from Table \ref{checktable}, for many sets of $\mathrm{\varphi_{1}}$, $\mathrm{\varphi_{2}}$; along with $\mathrm{m_{\beta \beta}}$ we can bring in 4 out of 5 oscillation parameters inside their 3-$\sigma$ ranges. One of the mass squared difference could not be brought inside its allowed range for any choice of the free parameters.
However, in past works it was shown that certain SUSY threshold corrections can have impact on the low scale values of the neutrino oscillation parameters, in particular the values of the mass square differences~\cite{Mohapatra:2005gs, Abbas:2014ala}. The important threshold correction relevant to our analysis are given in Appendix~\ref{app:threshold}.

\begin{figure}[!ht]
    \begin{subfigure}{.46 \textwidth}
      \includegraphics[width=\linewidth]{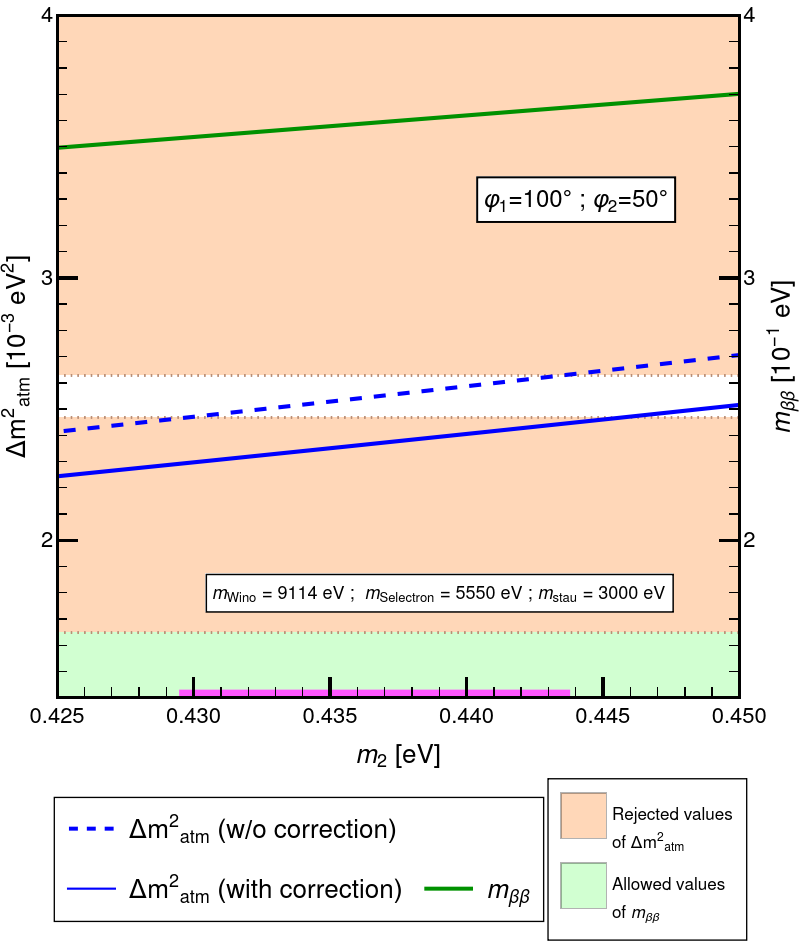}
      \caption{$\Delta \mathrm{m}^2_{\mathrm{atm}}$ and vs $\mathrm{m_{2}}$}
      \label{A_a}
    \end{subfigure}%
    \hspace{23 pt}
    \begin{subfigure}{.47\textwidth}
      \captionsetup{oneside,margin={0.8cm,0cm}}       \includegraphics[width=\linewidth]{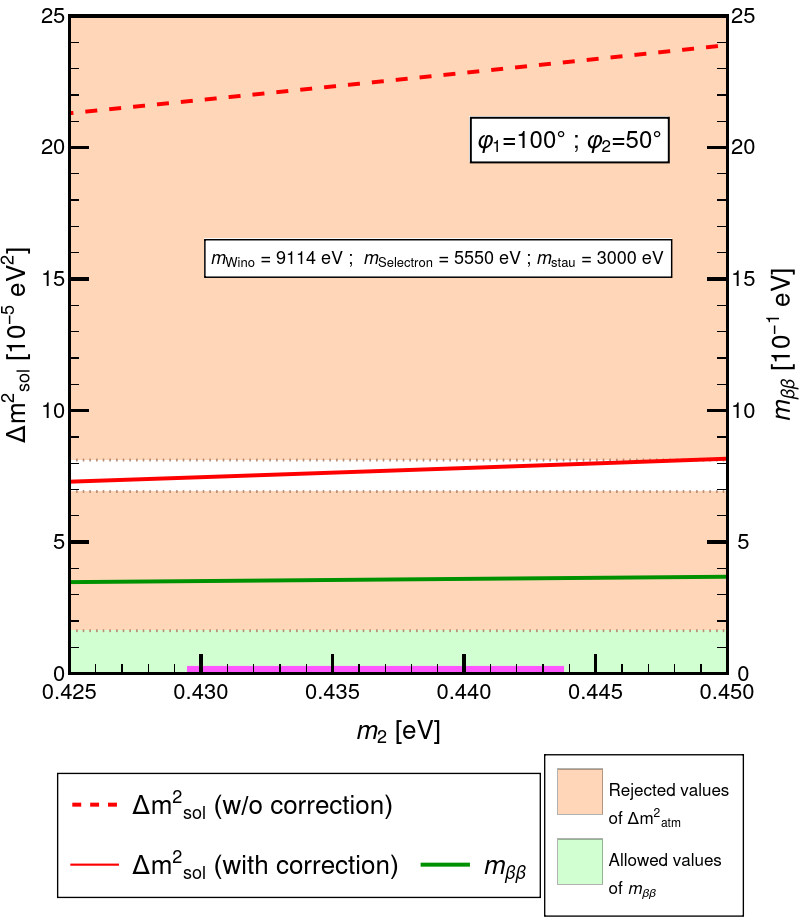}
      \caption{$\Delta \mathrm{m}^2_{\mathrm{sol}}$ and vs $\mathrm{m_{2}}$}
        \label{A_s}
    \end{subfigure}%
\caption{Impact of threshold corrections on $\Delta \mathrm{m}^2_{\mathrm{atm}}$ (left panel) and $\Delta \mathrm{m}^2_{\mathrm{sol}}$ vs $\mathrm{m_{2}}$ (right panel) with respect to $\mathrm{m_{2}}$ mass eigen-state.  The plots are done for the case where the uncorrected $\Delta \mathrm{m}^2_{\mathrm{atm}}$ was inside its 3-$\sigma$ range (white band) and $\Delta \mathrm{m}^2_{\mathrm{sol}}$ is brought inside its allowed range after threshold corrections. }
    \label{ATM}
\end{figure}

\begin{figure}[!ht]
    \begin{subfigure}{.46 \textwidth}
      \includegraphics[width=\linewidth]{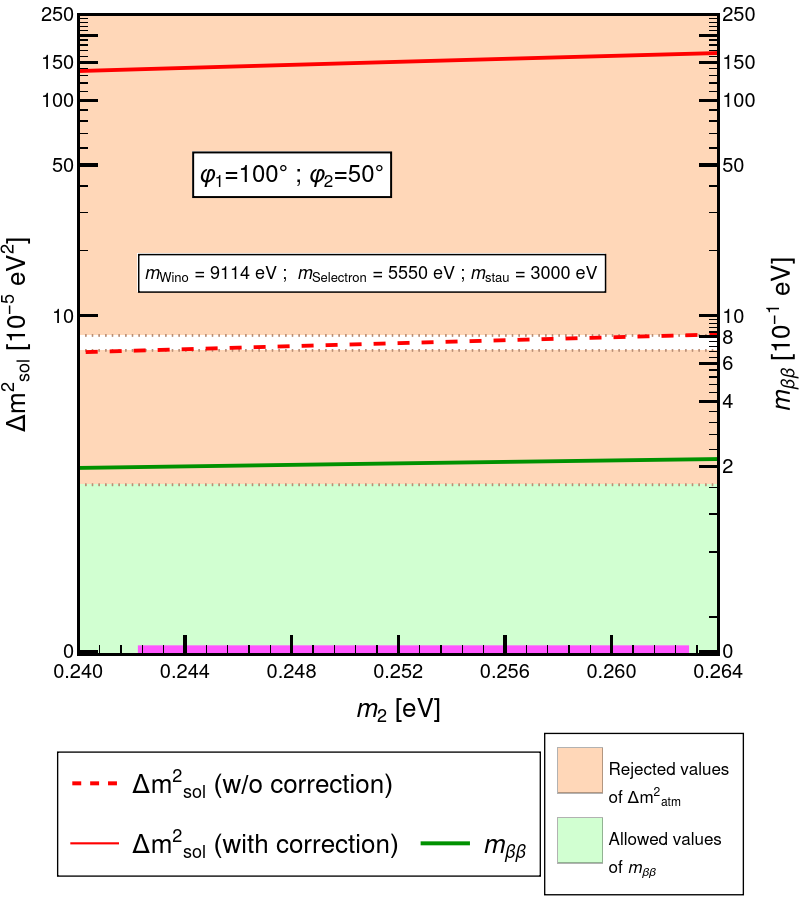}
      \caption{$\Delta \mathrm{m}^2_{\mathrm{sol}}$ and vs $\mathrm{m_{2}}$}
      \label{S_s}
    \end{subfigure}%
    \hspace{23 pt}
    \begin{subfigure}{.47\textwidth}
      \includegraphics[width=\linewidth]{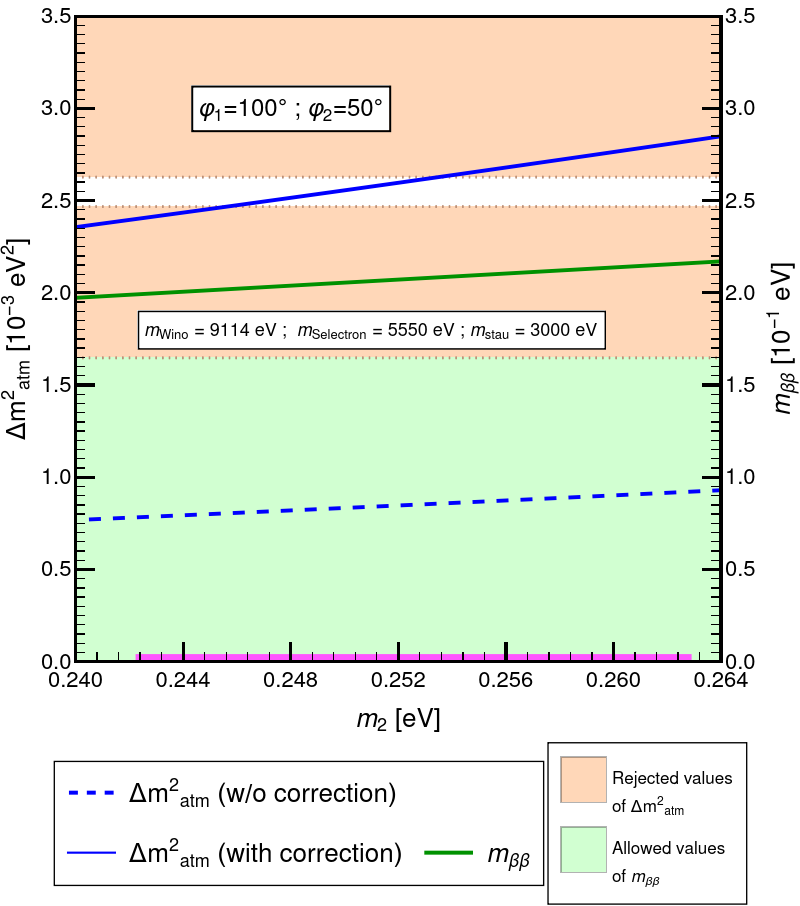}
      \caption{$\Delta \mathrm{m}^2_{\mathrm{atm}}$ and vs $\mathrm{m_{2}}$}
        \label{S_a}
    \end{subfigure}%
\caption{Impact of threshold corrections on $\Delta \mathrm{m}^2_{\mathrm{atm}}$ (left panel) and $\Delta \mathrm{m}^2_{\mathrm{sol}}$ vs $\mathrm{m_{2}}$ (right panel) with respect to $\mathrm{m_{2}}$ mass eigen state. This figure assumes that the uncorrected $\Delta \mathrm{m}^2_{\mathrm{sol}}$ is inside its range (white band) and $\Delta \mathrm{m}^2_{\mathrm{atm}}$ is brought inside its allowed range after threshold corrections.}
    \label{SOL}
\end{figure}

In order to see if inclusion of threshold corrections can lead to a viable parameter space consistent with current experimental data, we try to first see how that change the lower values of $\Delta \mathrm{m}^2_{\mathrm{atm}}$ and $\Delta \mathrm{m}^2_{\mathrm{sol}}$.
For this we choose one of the most promising pair of values of $\mathrm{\varphi_{1}}$ $= 100^\circ$ and $\mathrm{\varphi_{2}}$ $= 50^\circ$ from Tab.~\ref{checktable}, for which we were able to bring  one mass squared difference inside its 3-$\sigma$ range. 
After we add the threshold corrections large enough to bring the other mass square difference inside its 3-$\sigma$ range,  we find that the other mass squared difference which was already inside its allowed range, now moves out of the experimental 3-$\sigma$ range as shown in Figs.~\ref{ATM} and \ref{SOL}. 

Fig.\ref{ATM} is plotted for the case when $\Delta \mathrm{m}^2_{\mathrm{atm}}$ is inside its 3-$\sigma$ range without threshold corrections and $\Delta \mathrm{m}^2_{\mathrm{sol}}$ is outside. For this case we choose the masses etc of the sparticles such that the threshold corrections are large enough  to bring $\Delta \mathrm{m}^2_{\mathrm{sol}}$ inside its 3-$\sigma$ range, see eqs.~\eqref{solcorr} - \eqref{correctedm2} in Appendix~\ref{app:threshold}. 
After adding the corrections with appropriate parameters set, we find that the corrected $\Delta \mathrm{m}^2_{\mathrm{sol}}$ can indeed be brought inside its 3-$\sigma$ range, Fig.\ref{A_s}. We plot this against one of the GUT scale mass- $\mathrm{m_{2}}$. The corresponding threshold corrections will naturally be added in $\Delta \mathrm{m}^2_{\mathrm{atm}}$ too. But for the same set of $\mathrm{m_{2}}$ values for which corrected $\Delta \mathrm{m}^2_{\mathrm{sol}}$ was brought in; the corrected $\Delta \mathrm{m}^2_{\mathrm{atm}}$ is significantly far from its 3-$\sigma$ range. Fig.\ref{A_a}.

The same process is repeated in the Fig.\ref{SOL}; except now $\Delta \mathrm{m}^2_{\mathrm{sol}}$ is inside and $\Delta \mathrm{m}^2_{\mathrm{atm}}$ outside its range, before threshold corrections are added. Even in this case, for the same range of $\mathrm{m_{2}}$ values the threshold corrected $\Delta \mathrm{m}^2_{\mathrm{sol}}$ shifts outside its 3-$\sigma$ range, Fig.\ref{S_s} when the corrected $\Delta \mathrm{m}^2_{\mathrm{atm}}$  is successfully brought in its respective 3-$\sigma$ range, Fig.\ref{S_a}.


Thus in summary we have systematically analyzed all possible cases of HSMU ansatz, both for Dirac and Majorana neutrinos. Overall, we can conclude that HSMU ansatz is in conflict with the current 3-$\sigma$ allowed global fit ranges for neutrinos oscillation parameters and constraints from $0\nu \beta \beta$ decays in case of Majorana neutrinos.  Today's narrow 3-$\sigma$ experimental ranges of neutrino oscillation parameters do not allow HSMU hypothesis to be a plausible unification idea. In the next segment of the paper we discuss about an ansatz which generalizes the HSMU ansatz by imposing a looser set of demands on the high scale structure of the leptonic mixing matrix.

\section{Wolfenstein ansatz}
\label{wolf}

It was evident from previous sections that in no case the HSMU ansatz's prediction for all low scale neutrino parameters is consistent with their current 3-$\sigma$ ranges. For this reason, we will like to consider another ansatz which tries to explain the hierarchical nature of neutrino mixing angles~\cite{Abbas:2015vba,Abbas:2016qbl,AbdusSalam:2019hov,Rajasekaran:2019uvd}. This ansatz lifts the stringent restrictions on high scale leptonic mixing angles put by HSMU. In HSMU we consider that the leptonic mixing angles are exactly equal to those of quarks at the high scale (GUT scale). But instead, here we consider that the hierarchy in high scale quark mixing angle is duplicated in leptonic angles as well. This hierarchy is parameterized by ``Wolfenstein-like'' form. In this case the leptonic mixing angles are in the following pattern
\begin{equation}
\begin{split}
    \sin{\theta_{12}} = \lambda \hspace{1 cm}
    \sin{\theta_{23}} = \lambda^{2} \hspace{1 cm}
    \sin{\theta_{23}} = \lambda^{3}
\end{split}
\end{equation} 
where $\lambda$ is the leptonic Wolfenstein parameter which we define as the sine of the $\theta_{12}$ mixing angle.  Wolfenstein parameter gives the neutrino mixing angles the following hierarchical structure-
\begin{equation}
\begin{split}
    \theta_{12} = \arcsin(\lambda) \hspace{1 cm}
    \theta_{23} = \arcsin(\lambda^{2}) \hspace{1 cm}
    \theta_{23} = \arcsin(\lambda^{3})
\end{split}
\label{wolf_w/o_alphbet}
\end{equation}
We can vary $\theta_{23}$ and $\theta_{13}$ more finely by introducing $\mathrm{\alpha}$ and $\mathrm{\beta}$ such that $\alpha, \beta > 0$, Eq.~\eqref{wolf_w/o_alphbet} gets modified into 
\begin{equation}
\begin{split}
    \theta_{12} = \arcsin(\lambda) \hspace{1 cm}
    \theta_{23} = \alpha \arcsin(\lambda^{2}) \hspace{1 cm}
    \theta_{13} = \beta \arcsin(\lambda^{3})
\end{split}
\end{equation}
Note that, choosing $\lambda$ to be equal to $0.22532$, $\mathrm{\alpha}$ to be equal to $0.698353$ and $\mathrm{\beta}$ to be equal to $0.264066$ we get back the HSMU ansatz. 

Since from HSMU ansatz we already saw that only the case of Majorana neutrinos (with non-zero values of $\mathrm{\varphi_{1}}$, $\mathrm{\varphi_{2}}$) is close to being viable, therefore for Wolfenstein ansatz we will limit ourselves to this case only. We will follow the same strategy already discussed for HSMU but now with the relaxed condition of Wolfenstein ansatz on the high scale leptonic mixing angles.  
For the initial study of Wolfenstein ansatz, we analyze the effects of varying $\mathrm{\alpha}$ and $\mathrm{\beta}$ keeping Majorana phases to be zero and $\lambda$ equal to the value corresponding to the HSMU ansatz($\mathrm{\mathrm{\lambda_{HSMU}}} = 0.22532$). 
Once we find a suitable pair of $\mathrm{\alpha}$, $\mathrm{\beta}$, we can study variations in $\lambda$. Only after thoroughly analyzing effects of $\mathrm{\alpha}$, $\mathrm{\beta}$ and $\lambda$ we will bring in non-zero Majorana phases into the picture. The CP phase $\delta$ is taken to be equal to that of quarks CP violation phase($\mathrm{\delta_{q}}$) in all the cases.

\subsection{Variations of $\mathrm{\alpha}$ and $\mathrm{\beta}$} 

\begin{figure}[!ht]
    \centering
    \begin{subfigure}{0.45\textwidth}
        \centering
        \includegraphics[width=\linewidth]{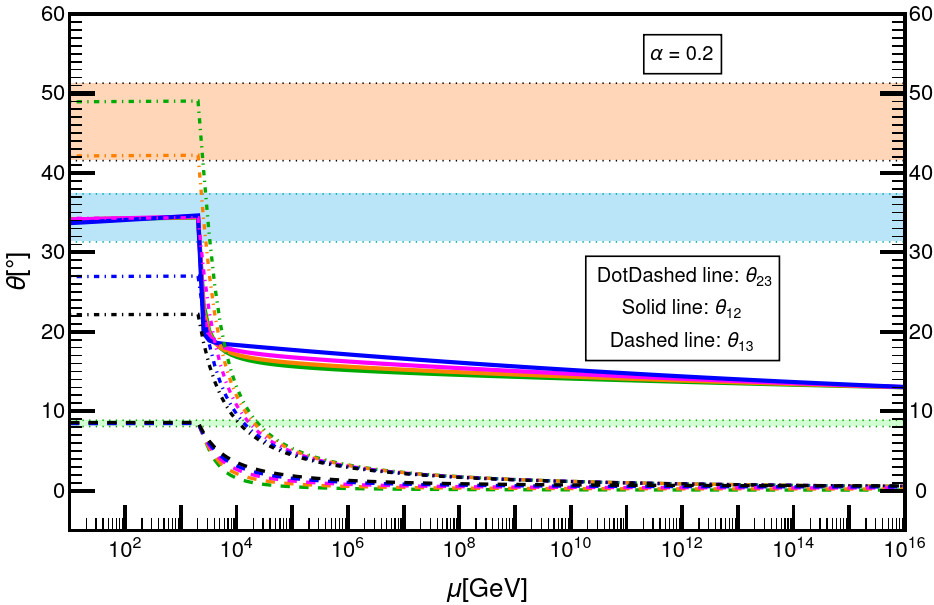}
        \caption{$\alpha = 0.2$}
        \label{wolf_rg_alph0.2}
    \end{subfigure}
    \begin{subfigure}{0.45\textwidth}
        \centering
        \includegraphics[width=\linewidth]{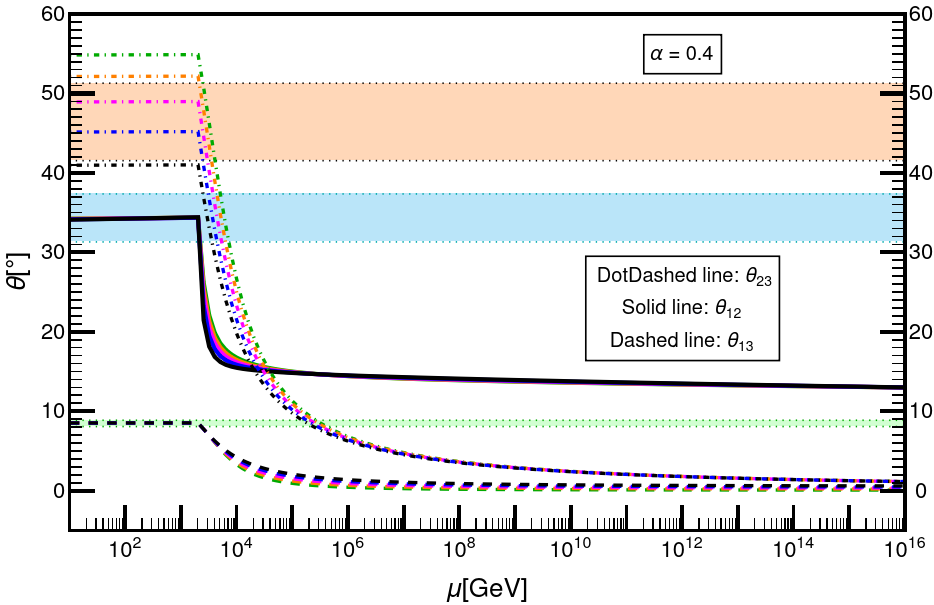}
        \caption{$\alpha = 0.4$}
        \label{wolf_rg_alph0.4}
    \end{subfigure} 
    \begin{subfigure}{0.45\textwidth}
        \centering
        \includegraphics[width=\linewidth]{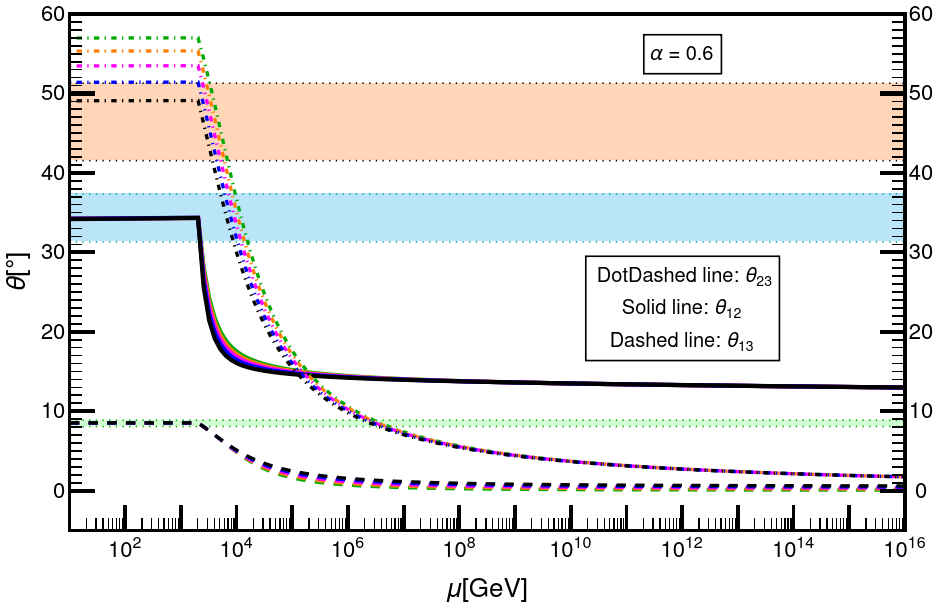}
        \caption{$\alpha = 0.6$}
        \label{wolf_rg_alph0.6}
    \end{subfigure} 
    \begin{subfigure}{.45\textwidth}
        \centering
        \includegraphics[width=\linewidth]{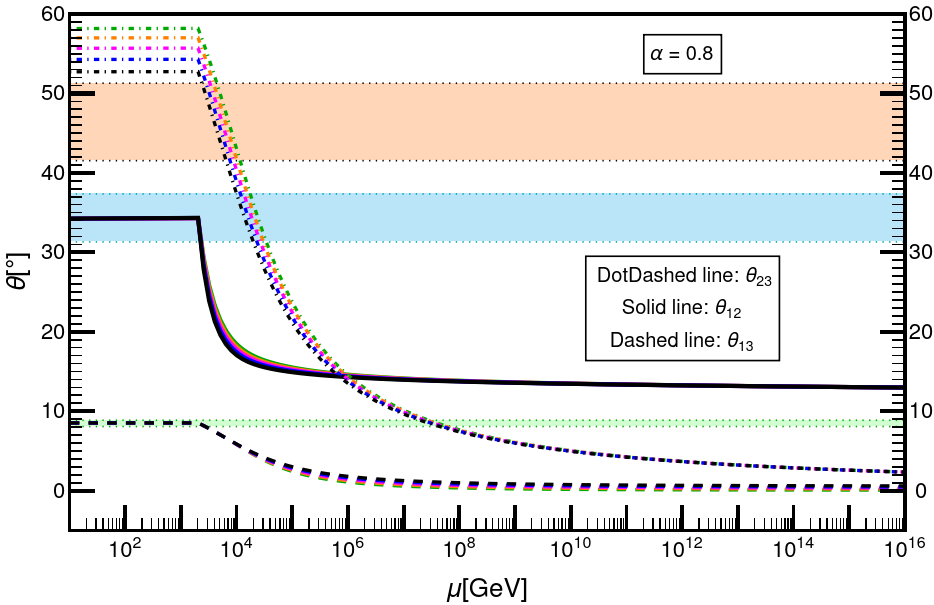}
        \caption{$\alpha = 0.8$}
        \label{wolf_rg_alph0.8}
    \end{subfigure}  
    \begin{subfigure}{0.45\textwidth}
        \centering
        \includegraphics[width=1.3\linewidth]{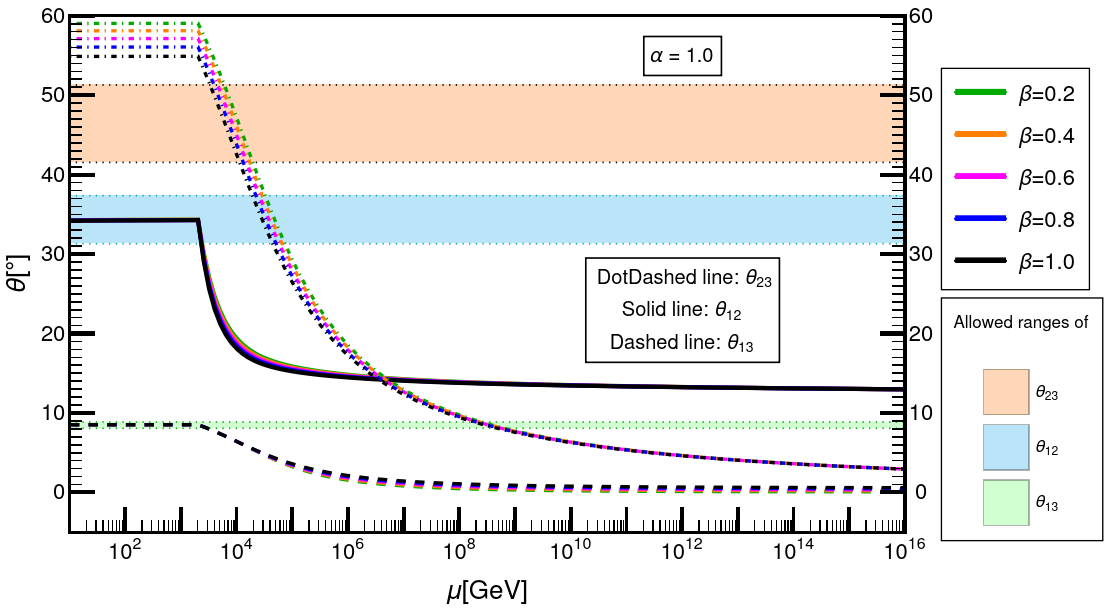}
        \caption{$\alpha = 1.0$}
        \label{wolf_rg_alph1.0}
    \end{subfigure}
    \caption[]{The RG evolution of neutrino mixing angles for various $\mathrm{\alpha}$, $\mathrm{\beta}$ pairs. Value of $\mathrm{\alpha}$ is kept fixed for every graph and $\mathrm{\beta}$ is varied, represented by different colours. The shaded regions show 3-$\sigma$ ranges for all three angles. For all the graphs $\lambda = \mathrm{\lambda_{HSMU}} = 0.22532$.}
    \label{wolf_rg}
\end{figure}

Changing $\mathrm{\alpha}$ and $\mathrm{\beta}$ can change the hierarchy of the angles at high scale which can possibly alter the usual RG running of the angles. Hence, to study variations of $\mathrm{\alpha}$, $\mathrm{\beta}$ we first analyze how the angles RG evolve for each pair of $\mathrm{\alpha}$, $\mathrm{\beta}$ we choose.

To obtain the graphs in Fig.~\ref{wolf_rg} neutrino masses at high scales are chosen such that $\theta_{12}$ and $\theta_{13}$ are at their best fit value at low scale and $\theta_{23}$ at low scale is allowed to vary. The trend across the graphs shows that as $\mathrm{\alpha}$ increases the span of $\theta_{23}$ shrinks and shifts higher in values. As a result of this, for a particular value of $\mathrm{\alpha}$, only for a few  $\mathrm{\beta}$ values, all three angles have values within their experimental 3-$\sigma$ range. For example from Fig.~\ref{wolf_rg_alph0.8} and Fig.~\ref{wolf_rg_alph1.0} it can be seen that the entire span of $\theta_{23}$ lies outside of its valid range, i.e. for $\alpha = 0.8$ and $\alpha = 1.0$, we can not have all three mixing angles inside their valid ranges.


\subsection{Variation of $\lambda$}


\begin{figure}
\vspace{-0.7 cm}
 \begin{subfigure}{0.45\textwidth} \hspace{-70 pt}
  \centering
  \includegraphics[width=1\linewidth]{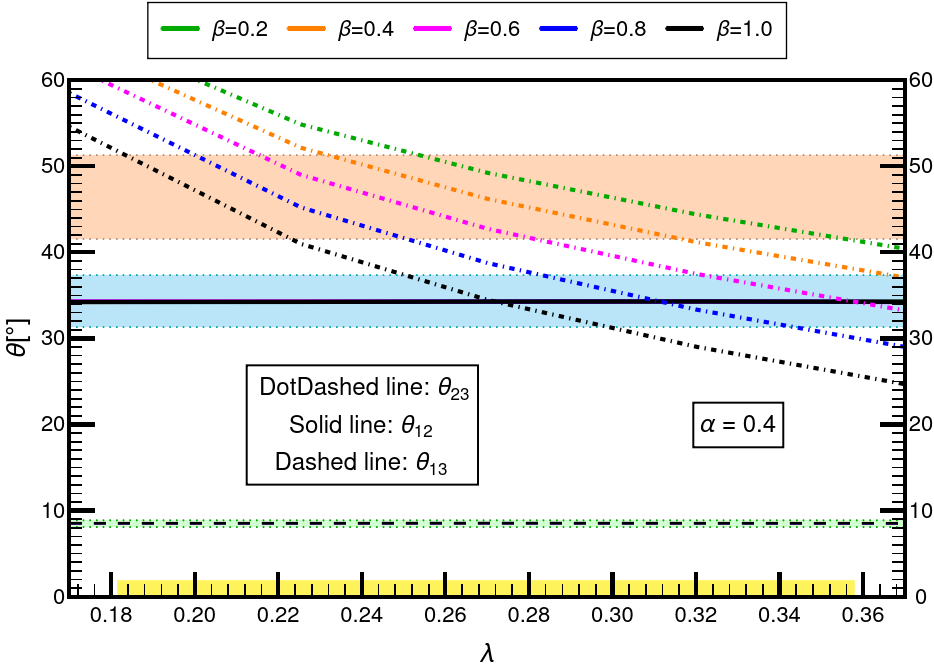}
  \caption{$\theta_{12}$, $\theta_{23}$, $\theta_{13}$ vs $\lambda$ for $\alpha = 0.4$}
  \label{lamvsang_0.4}
\end{subfigure}
\begin{subfigure}{0.474\textwidth} \vspace{5.5 pt}
  \centering
  \vspace{-7pt}
  \includegraphics[width=1\linewidth]{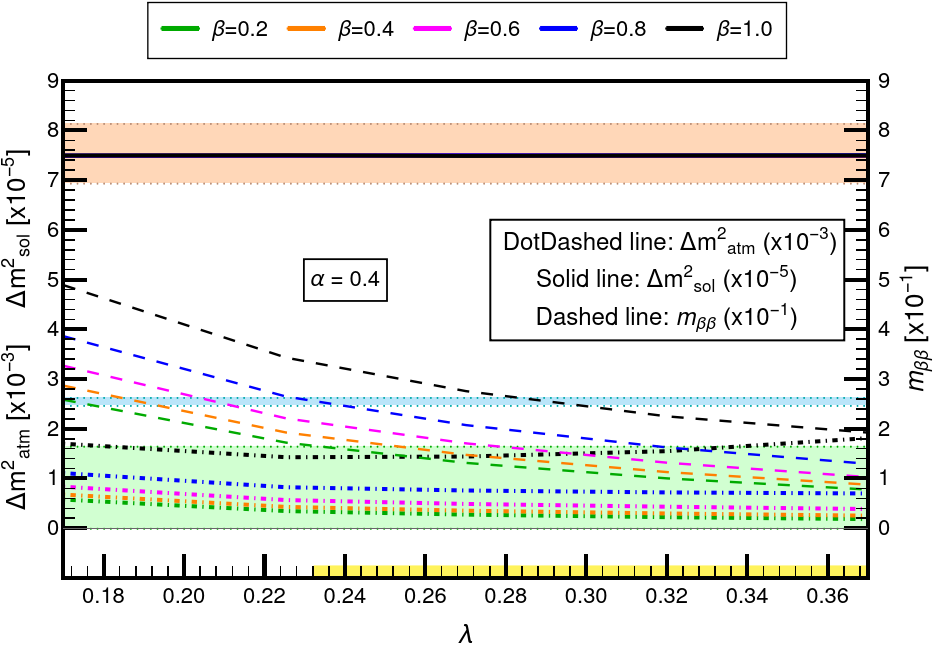}
  \caption{$\Delta \mathrm{m^{2}_{sol}}$, $\Delta \mathrm{m^{2}_{atm}}$, $\mathrm{m_{\beta \beta}}$ vs $\lambda$ for $\alpha = 0.4$}
  \label{lamvsmass_0.4}
\end{subfigure} \\ \vspace{15 pt}
\begin{subfigure}{0.45\textwidth} \hspace{-70 pt}
  \centering
  \includegraphics[width=1\linewidth]{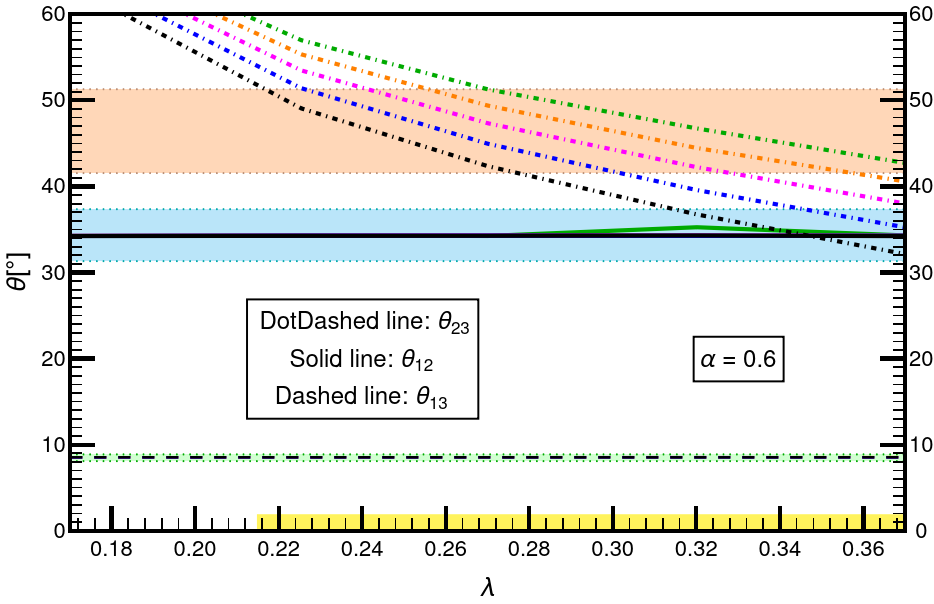}
  \caption{$\theta_{12}$, $\theta_{23}$, $\theta_{13}$ vs $\lambda$ for $\alpha = 0.6$}
  \label{lamvsang_0.6}
\end{subfigure}
\begin{subfigure}{0.474\textwidth} \vspace{5.5 pt}
  \centering
  \vspace{-7pt}
  \includegraphics[width=1\linewidth]{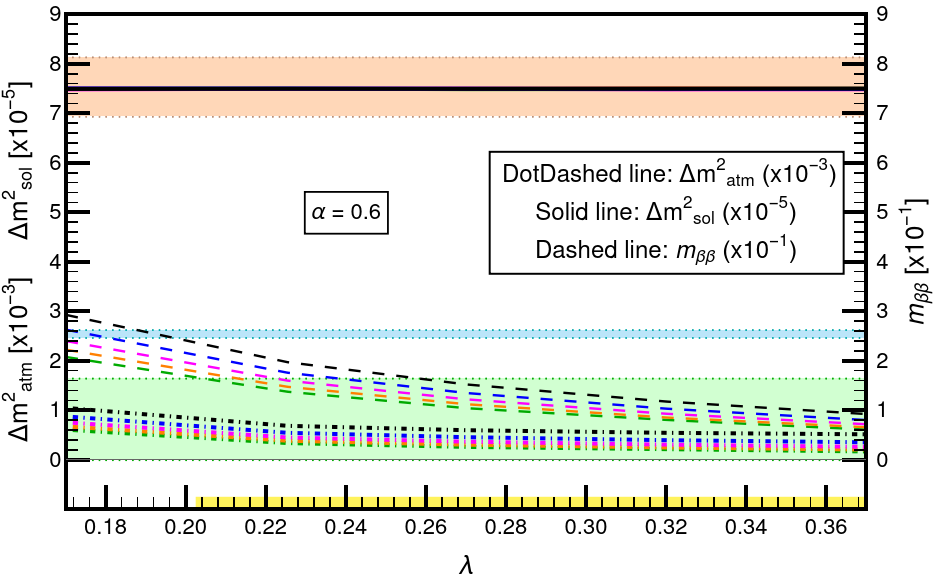}
  \caption{$\Delta \mathrm{m^{2}_{sol}}$, $\Delta \mathrm{m^{2}_{atm}}$, $\mathrm{m_{\beta \beta}}$ vs $\lambda$ for $\alpha = 0.6$}
  \label{lamvsmass_0.6}
\end{subfigure} 
\label{lamvary_1}
%
%
%
%
\begin{subfigure}[!ht]{0.45\textwidth} \hspace{-70 pt}
  \centering
  \includegraphics[width=1\linewidth]{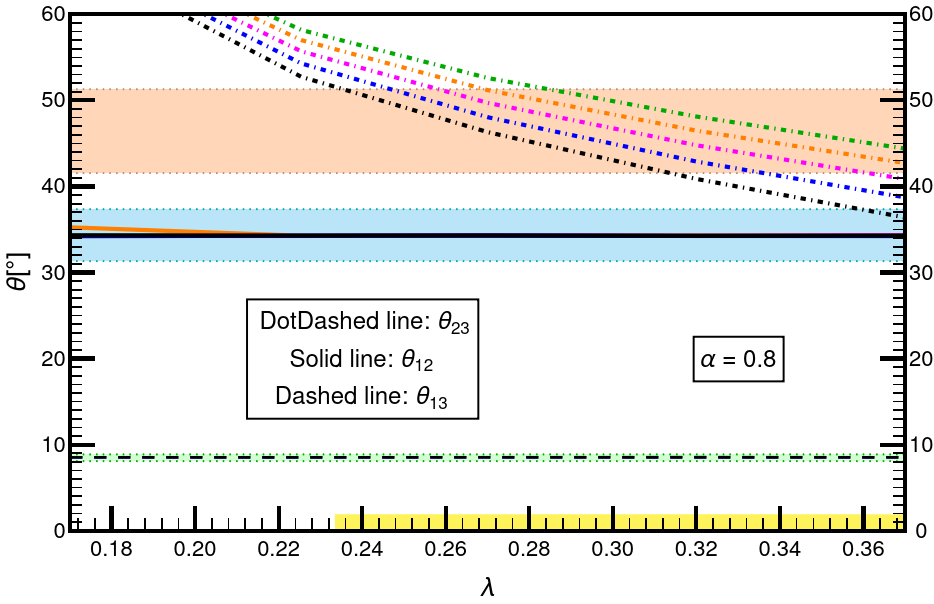}
  \caption{ $\theta_{12}$, $\theta_{23}$, $\theta_{13}$ vs $\lambda$ for $\alpha = 0.8$}
  \label{lamvsang_0.8}
\end{subfigure}
\begin{subfigure}{0.474\textwidth} \vspace{5.5 pt}
  \centering
  \vspace{-7pt}
  \includegraphics[width=1\linewidth]{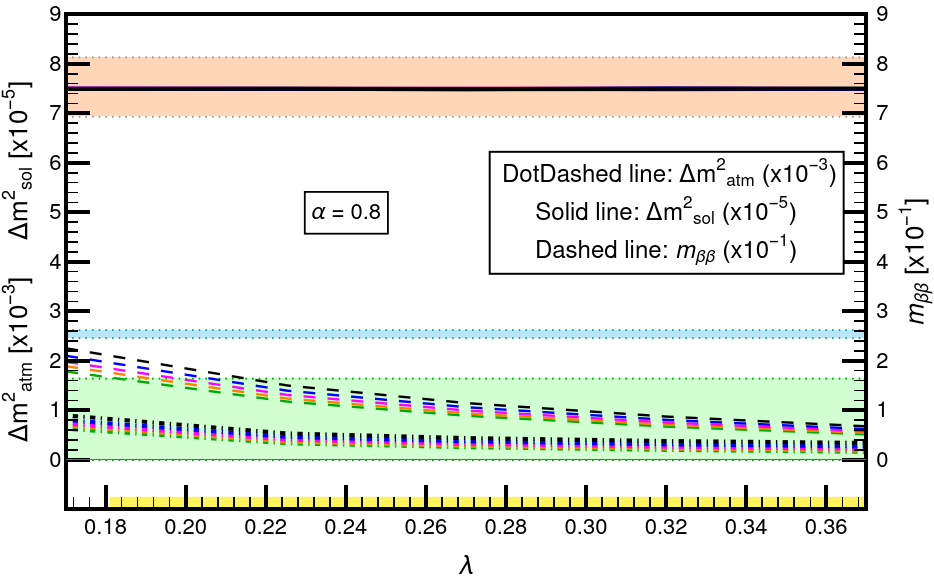}
  \caption{$\Delta \mathrm{m^{2}_{sol}}$, $\Delta \mathrm{m^{2}_{atm}}$, $\mathrm{m_{\beta \beta}}$ vs $\lambda$ for $\alpha = 0.8$}
  \label{lamvsmass_0.8}
\end{subfigure} \\

\begin{subfigure}{0.45\textwidth} \hspace{-70 pt}
  \centering
  \includegraphics[width=1\linewidth]{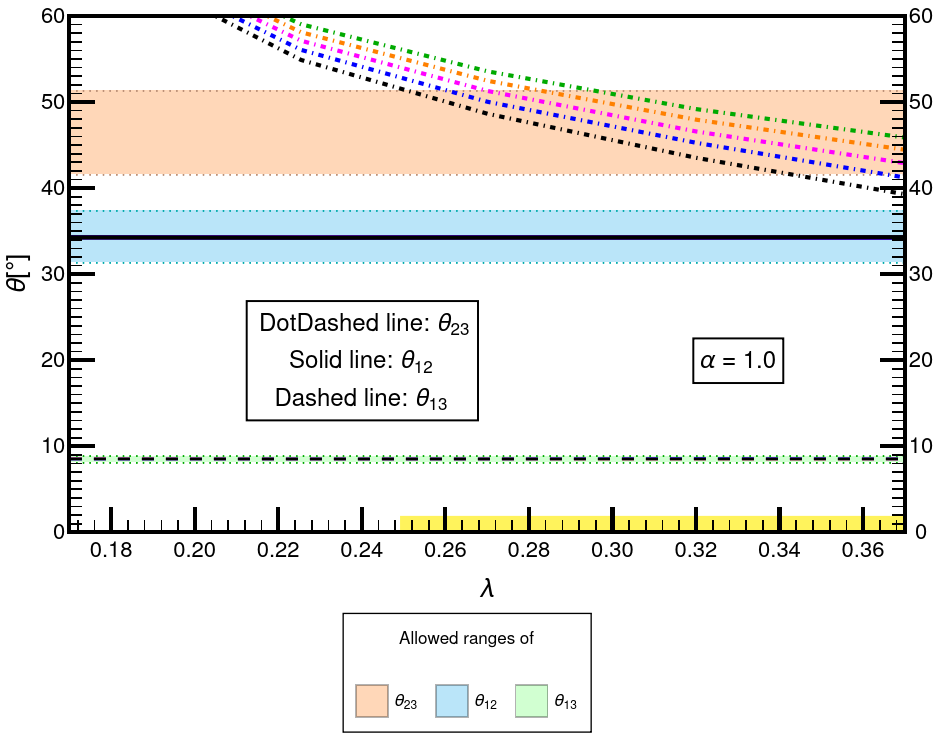}
  \caption{$\theta_{12}$, $\theta_{23}$, $\theta_{13}$ vs $\lambda$ for $\alpha = 1.0$}
  \label{lamvsang_1.0}
\end{subfigure}
\begin{subfigure}{0.474\textwidth} \vspace{5.5 pt}
  \centering
  \vspace{-7pt}
  \includegraphics[width=1\linewidth]{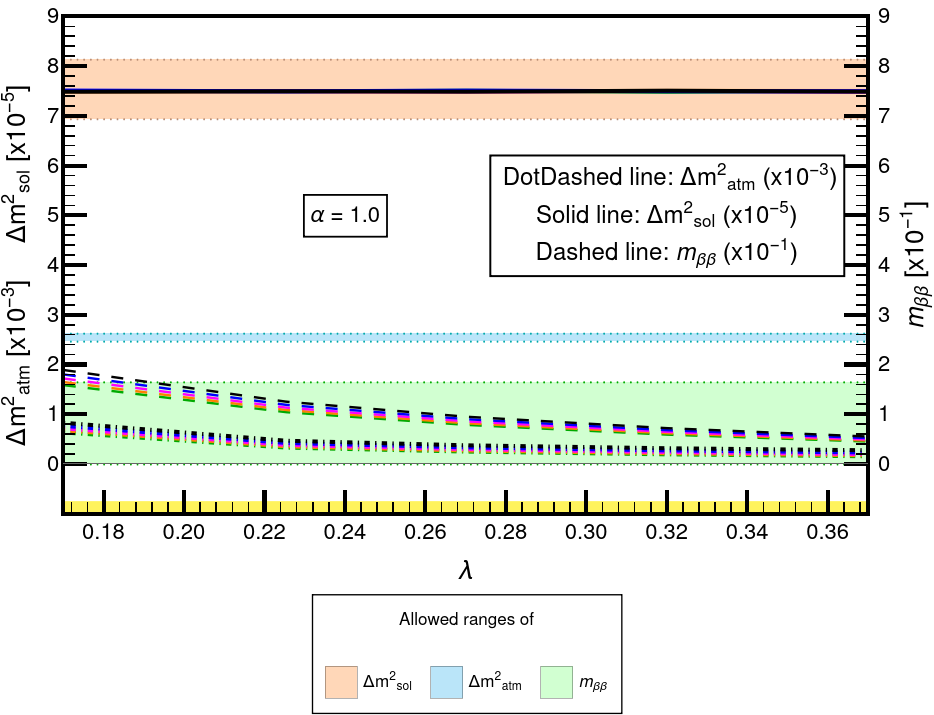}
  \caption{$\Delta \mathrm{m^{2}_{sol}}$, $\Delta \mathrm{m^{2}_{atm}}$, $\mathrm{m_{\beta \beta}}$ vs $\lambda$ for $\alpha = 1.0$}
  \label{lamvsmass_1.0}
\end{subfigure}

\caption{\begin{footnotesize}  Impact of the variation of $\lambda$ on all the low scale parameter. For each graph, value of $\mathrm{\alpha}$ is kept fixed and $\mathrm{\beta}$ is varied. The shaded regions show 3-$\sigma$ ranges for all three angles, mass square differences and $\mathrm{m_{\beta \beta}}$. See text for more details.                                                                                                                                                                                                                                                                                                                                             \end{footnotesize}}

\label{lamvariations}
\end{figure}

Having understood how different values of $\alpha, \beta$ are changing the RG evolution of the mixing angles, let us now see how variation in $\lambda$ values changes it. Out of 25 pairs of $\mathrm{\alpha}$, $\mathrm{\beta}$ analyzed in previous section we choose the 20 valid pairs for which we analyze low scale values of $\theta_{12}$, $\theta_{23}$, $\theta_{13}$, $\Delta \mathrm{m^{2}_{sol}}$, $\Delta \mathrm{m^{2}_{atm}}$ as well as $\mathrm{m_{\beta \beta}}$. Note that we have still kept the  Majorana phases to be zero. Its effects are to be seen later.

Fig.~\ref{lamvariations} shows the impact of variation of $\lambda$ on all the low scale  mixing parameters as well as  $\mathrm{m_{\beta \beta}}$.  We use high scale neutrino masses (free parameters) such that $\theta_{12}$, $\theta_{13}$ and $\Delta \mathrm{m^{2}_{sol}}$ are fixed at their best fit values. Since, all the parameters are correlated, the RG evolution of  $\theta_{23}$, $\Delta \mathrm{m^{2}_{atm}}$ and $\mathrm{m_{\beta \beta}}$ is allowed to vary freely dictated the by the RG equations. 

As we can see $\theta_{23}$ and $\mathrm{m_{\beta \beta}}$ values decrease for increasing $\lambda$. While doing so, they cross their valid range. The range of $\lambda$ values where for some choice of other free parameters, at least one $\lambda$ value leads to either $\theta_{23}$ or $\mathrm{m_{\beta \beta}}$ inside its allowed range, is highlighted on X-axis of Fig.\ref{lamvariations} by yellow color. For a specific pair of $\mathrm{\alpha}$ and $\mathrm{\beta}$ we can compare these minimum values of $\lambda$ above which all angles and $\mathrm{m_{\beta \beta}}$ can be brought inside their respective 3-$\sigma$ ranges. For example, in $\alpha = 0.2$ case, the $\lambda$ -space for which all angles are inside allowed ranges does not overlap with $\lambda$ -space for which $\mathrm{m_{\beta \beta}}$ is inside its 3-$\sigma$ range. Therefore, that case is omitted in Fig.~\ref{lamvariations}. Although, note that even for $\alpha = 0.2$, we can still bring four out of the six parameters inside their allowed ranges, which is still an improvement over the HSMU case, where we could not even bring all three angles inside their range. This is an indication that Wolfenstein ansatz can potentially be more promising.

For next set of $\mathrm{\alpha}$ values shown in Fig.~\ref{lamvariations}, the results are even more promising. We can find an overlap between $\lambda$ -spaces of angles (left panels)  and $\mathrm{m_{\beta \beta}}$  (right panels) for same $\mathrm{\alpha}$. It means that for those $\lambda$ values, we can bring in all angles, $\Delta \mathrm{m^{2}_{sol}}$ as well as $\mathrm{m_{\beta \beta}}$ inside their 3-$\sigma$ ranges, i.e. five out of six parameters but never all six simultaneously.
Thus we have to investigate further by taking non-zero values of the  Majorana phases ($\mathrm{\varphi}_1$, $\mathrm{\varphi}_2$) and study its effects on all the low scale parameters and see if all six can be brought inside.

\subsection{$\mathrm{\varphi}_1$, $\mathrm{\varphi}_2$ variations}

Here we will not span the entire $\mathrm{\varphi}_1$, $\mathrm{\varphi}_2$ space unnecessarily as it is computationally very expensive. In any case, the final goal is to find a set of input parameters for which all six parameters (angles, mass squared differences, $\mathrm{m_{\beta \beta}}$) are inside their 3-$\sigma$ ranges at low scale. 
We focus on the parameters which are set free to vary, i.e., $\theta_{23}$ , $\Delta \mathrm{m^{2}_{atm}}$ and $\mathrm{m_{\beta \beta}}$. This is because other parameters are held constant at their best fit values using the three independent parameters namely the high scale neutrino masses. Thus, the newly introduced parameters $\lambda$ , $\mathrm{\alpha}$ and $\mathrm{\beta}$ along with Majorana phases can be scanned to find possible ranges where the  low scale values of $\theta_{23}$ , $\Delta \mathrm{m^{2}_{atm}}$ and $\mathrm{m_{\beta \beta}}$ are within their 3-$\sigma$ range.

To have a simple and convenient start, we use the HSMU case to pick up only those $\mathrm{\varphi}_1$, $\mathrm{\varphi}_2$ pairs for which we could bring in all angles and one of the mass squared differences inside valid ranges. In Table.\ref{checktable}, we list the values of $\mathrm{\varphi}_1$ , $\mathrm{\varphi}_2$  where four oscillation parameters can be brought inside their 3-$\sigma$ range.
\begin{table}[!ht]
 \centering
\begin{tabular}{ |p{30 pt}|p{30 pt}|p{20 pt}|p{20 pt}|p{20 pt}|p{35 pt}|p{40 pt}| }  
 \hline
 $\varphi_{1}(^{\circ})$ & $\varphi_{2}(^{\circ})$ & $\theta_{12} $ & $\theta_{13} $ & $\theta_{23} $ & $\Delta m^{2}_{sol} $ & $\Delta m^{2}_{atm} $\\ \hline \hline
 
 50 & 0 & \checkmark & \checkmark  & \checkmark  & \multicolumn{2}{c|}{Only one}  \\ \hline 

 50 & 300 & \checkmark & \checkmark  & \checkmark  & \multicolumn{2}{c|}{Only one}  \\ \hline 

 200 & 300 & \checkmark & \checkmark  & \checkmark  & \multicolumn{2}{c|}{Only one}  \\ \hline 

 300 & 50 & \checkmark & \checkmark  & \checkmark  & \multicolumn{2}{c|}{Only one}  \\ \hline 
 \end{tabular}
\end{table}

For these values, we then discretely vary $\mathrm{\alpha}$ and $\mathrm{\beta}$ . From $\mathrm{\alpha}$, $\mathrm{\beta}$ and $\lambda$ variations shown in Fig.(\ref{lamvariations}), it is observed that as $\lambda$ increases, $\theta_{23}$ , $\Delta \mathrm{m^{2}_{atm}}$ and $\mathrm{m_{\beta \beta}}$ decrease and vice versa.
An exception to this trend is observed at lower values of $\mathrm{\alpha}$, where $\Delta \mathrm{m^{2}_{atm}}$ behaves the opposite for larger values of $\lambda$. One more reason to ignore low $\mathrm{\alpha}$ values is that the span of $\theta_{23}$ is larger than its 3-$\sigma$ range. Because of this, all $\mathrm{\beta}$ values for $\alpha=0.2$ and $\alpha=0.4$ should be discarded as for these values one can not bring $\theta_{23}$ value inside its 3-$\sigma$ range.

Keeping this in mind, we ignore the lower $\mathrm{\alpha}$ values here. 
For now, we shall only vary $\lambda$ for those pairs of $\mathrm{\varphi}_1$ , $\mathrm{\varphi}_2$ for which $\theta_{23}$ , $\Delta \mathrm{m^{2}_{atm}}$ and $\mathrm{m_{\beta \beta}}$ are all either above/below their 3-$\sigma$ ranges simultaneously or are all inside their 3-$\sigma$ ranges simultaneously. This is because if one of the parameters from $\theta_{23}$ , $\Delta \mathrm{m^{2}_{atm}}$ and $\mathrm{m_{\beta \beta}}$ lies on one side(higher or lower) of its own 3-$\sigma$ range and other parameter lies on the opposite side of its own 3-$\sigma$ range, changing $\lambda$ will result in shifting one of the parameters away from its 3-$\sigma$ range and thus not letting us bring in all the six low scale parameters in their valid experimental bounds. This plan is case-wise realised in the following sub-section.


\subsection{$\mathrm{\alpha}$, $\mathrm{\beta}$, $\lambda$, $\mathrm{\varphi}_1$, $\mathrm{\varphi}_2$ variations combined}

Now we shall choose the aforementioned $\mathrm{\varphi}_1$, $\mathrm{\varphi}_2$ pairs for $\mathrm{\alpha}$, $\mathrm{\beta}$ variation and select only those values for $\lambda$ variations, for which $\theta_{23}$, $\Delta \mathrm{m^{2}_{atm}}$ and $\mathrm{m_{\beta \beta}}$ lie on one side of their respective 3-$\sigma$ range. To illustrate this, a symbolic notation is used in order to understand the increasing/decreasing trend of these parameters.

Using various colours and shapes, Tables \ref{alphebet_symtable_1} and \ref{alphebet_symtable_2}  illustrate whether the respective parameter is inside, above or below the respective 3-$\sigma$ ranges for various pairs of $\mathrm{\alpha}$ , $\mathrm{\beta}$ and $\mathrm{\varphi}_1$, $\mathrm{\varphi}_2$. As discussed above, we ignore the lower values of $\mathrm{\alpha}$ ($0.4$ and $0.2$) as the $\lambda$ variations don't assure either monotonic increase or monotonic decrease in $\Delta \mathrm{m^{2}_{atm}}$, for these $\mathrm{\alpha}$ values. So in the rest, we want to get rid of those $\mathrm{\alpha}$, $\mathrm{\beta}$ pairs for which any two low scale parameters lie on opposite sides of their respective 3-$\sigma$ ranges. In the symbolic representation, this translates to ruling out those data sets which have both green and red coloured symbols for the same $\mathrm{\alpha}$, $\mathrm{\beta}$ pairs. Let's consider the four cases one by one.

\begin{table}[!ht]
    \centering
\begin{subtable}{.5\textwidth} \centering
\begin{tabular}{|c|c|c|c|c|} 
        \hline
$\mathrm{\alpha}$ & $\mathrm{\beta}$ &\hspace{0.1cm} $\theta_{23}$ \hspace{0.1cm}& \hspace{0.1cm} $\Delta \mathrm{m^{2}_{atm}}$ \hspace{0.1cm}&\hspace{0.1cm} $\mathrm{m_{\beta \beta}}$ \hspace{0.1cm} \\ \hline
\hspace{0.1cm} 1.0 \hspace{0.1cm}&\hspace{0.1cm} 1.0 \hspace{0.1cm} &\hspace{0.1cm} \textcolor{black}{$\bullet $}\hspace{0.1cm} &  
\hspace{0.1cm} \textcolor{red}{$\bullet $} \hspace{0.1cm} &  \hspace{0.1cm} \textcolor{grn}{$\bullet $} \hspace{0.1cm}\\ \hline
        
        1.0 & 0.8 & \textcolor{black}{$\star $} &  
        \textcolor{red}{$\bullet $} &  
        \textcolor{grn}{$\star $} \\ \hline
        
        1.0 & 0.6 & \textcolor{grn}{$\star $} &  
        \textcolor{red}{$\bullet $} & 
        \textcolor{black}{$\bullet $} \\ \hline
        
        1.0 & 0.4 & \textcolor{grn}{$\bullet $} &  
        \textcolor{red}{$\bullet $} &  
        \textcolor{black}{$\bullet $} \\ \hline
        
        1.0 & 0.2 & \textcolor{grn}{$\bullet $} &  
        \textcolor{red}{$\bullet $} &  
        \textcolor{black}{$\bullet $} \\ \hline \hline
        0.8 & 1.0 & \textcolor{red}{$\star $} &  
        \textcolor{grn}{$\bullet $} &  
        \textcolor{grn}{$\bullet $} \\ \hline
        
        0.8 & 0.8 & \textcolor{black}{$\bullet $} &  
        \textcolor{red}{$\bullet $} &  
        \textcolor{grn}{$\bullet $} \\ \hline
        
        0.8 & 0.6 & \textcolor{black}{$\star $} &  
        \textcolor{red}{$\bullet $} & 
        \textcolor{grn}{$\bullet $} \\ \hline
        
        0.8 & 0.4 & \textcolor{grn}{$\star $} &  
        \textcolor{red}{$\bullet $} &  
        \textcolor{black}{$\bullet $} \\ \hline
        
        0.8 & 0.2 & \textcolor{grn}{$\bullet $} &  
        \textcolor{red}{$\bullet $} &  
        \textcolor{black}{$\bullet $} \\ \hline \hline
        0.6 & 1.0 & \textcolor{red}{$\bullet $} &  
        \textcolor{grn}{$\bullet $} &  
        \textcolor{grn}{$\bullet $} \\ \hline
        
        0.6 & 0.8 & \textcolor{red}{$\star $} &  
        \textcolor{grn}{$\bullet $} &  
        \textcolor{grn}{$\bullet $} \\ \hline
        
        0.6 & 0.6 & \textcolor{black}{$\bullet $} &  
        \textcolor{red}{$\bullet $} &  
        \textcolor{grn}{$\bullet $} \\ \hline
        
        0.6 & 0.4 & \textcolor{black}{$\star $} &  
        \textcolor{grn}{$\bullet $} &  
        \textcolor{grn}{$\bullet $} \\ \hline
        
        0.6 & 0.2 & \textcolor{grn}{$\star $} &  
        \textcolor{red}{$\bullet $} &  
        \textcolor{black}{$\bullet $} \\ \hline \hline
        0.4 & 1.0 & \textcolor{red}{$\bullet $} &  
        \textcolor{red}{$\bullet $} &  
        \textcolor{grn}{$\bullet $} \\ \hline
        
        0.4 & 0.8 & \textcolor{red}{$\bullet $} &  
        \textcolor{grn}{$\bullet $} &  
        \textcolor{grn}{$\bullet $} \\ \hline
        
        0.4 & 0.6 & \textcolor{red}{$\bullet $} &  
        \textcolor{grn}{$\bullet $} &  
        \textcolor{grn}{$\bullet $} \\ \hline
        
        0.4 & 0.4 & \textcolor{black}{$\star $} &  
        \textcolor{red}{$\bullet $} &  
        \textcolor{grn}{$\bullet $} \\ \hline
        
        0.4 & 0.2 & \textcolor{black}{$\star $}  &  
        \textcolor{red}{$\bullet $} &  
        \textcolor{grn}{$\bullet $} \\ \hline \hline
        0.2 & 1.0 & \textcolor{red}{$\bullet $} &  
        \textcolor{grn}{$\bullet $} &  
        \textcolor{grn}{$\bullet $} \\ \hline
        
        0.2 & 0.8 & \textcolor{red}{$\bullet $} &  
        \textcolor{grn}{$\bullet $} &  
        \textcolor{grn}{$\bullet $} \\ \hline
        
        0.2 & 0.6 & \textcolor{red}{$\bullet $} &  
        \textcolor{grn}{$\bullet $} &  
        \textcolor{grn}{$\bullet $} \\ \hline
        
        0.2 & 0.4 & \textcolor{red}{$\bullet $} &  
        \textcolor{grn}{$\bullet $} &  
        \textcolor{grn}{$\bullet $} \\ \hline
        
        0.2 & 0.2 &\textcolor{red}{$\star $} &  
        \textcolor{red}{$\bullet $} &  
        \textcolor{grn}{$\bullet $} \\ \hline
        
    \end{tabular}
    \caption{\textbf{\hspace{1cm}For $\varphi_{1} = 50^{\circ}$ and $\varphi_{2} = 0^{\circ}$}}
    \label{alphebet_symtable_50_0}
    \end{subtable}%
    \begin{subtable}{.5\textwidth} \centering
        \begin{tabular}{|c|c|c|c|c|} \hline
        $\mathrm{\alpha}$ & $\mathrm{\beta}$ &\hspace{0.1cm} $\theta_{23}$ \hspace{0.1cm}&\hspace{0.1cm} $\Delta \mathrm{m^{2}_{atm}}$ \hspace{0.1cm}&\hspace{0.1cm} $\mathrm{m_{\beta \beta}}$ \hspace{0.1cm} \\ \hline
        \hspace{0.1cm} 1.0 \hspace{0.1cm}& \hspace{0.1cm} 1.0 \hspace{0.1cm}& \textcolor{black}{$\bullet $} &  \textcolor{red}{$\bullet $} &  \textcolor{grn}{$\bullet $} \\ \hline
        
        1.0 & 0.8 & $\star $ &  \textcolor{red}{$\bullet $} &  \textcolor{black}{$\bullet $} \\ \hline
        
        1.0 & 0.6 & $\star $ & \textcolor{red}{$\bullet $} &  \textcolor{black}{$\bullet $} \\ \hline
        
        1.0 & 0.4 & \textcolor{grn}{$\star $} & \textcolor{red}{$\bullet $} &  \textcolor{black}{$\bullet $} \\ \hline
        
        1.0 & 0.2 & \textcolor{grn}{$\star $} & \textcolor{red}{$\bullet $} &  \textcolor{black}{$\bullet $} \\ \hline \hline
        0.8 & 1.0 & \textcolor{black}{$\star $} &  \textcolor{grn}{$\bullet $} &  \textcolor{grn}{$\bullet $} \\ \hline
        
        0.8 & 0.8 & \textcolor{black}{$\bullet $} & \textcolor{red}{$\bullet $} &  \textcolor{grn}{$\bullet $} \\ \hline
        
        0.8 & 0.6 & \textcolor{black}{$\bullet $} & \textcolor{red}{$\bullet $} &  \textcolor{grn}{$\star $} \\ \hline
        
        0.8 & 0.4 & \textcolor{black}{$\star $} & \textcolor{red}{$\bullet $} &  \textcolor{black}{$\bullet $} \\ \hline
        
        0.8 & 0.2 & \textcolor{grn}{$\star $} & \textcolor{red}{$\bullet $} &  \textcolor{black}{$\bullet $} \\ \hline \hline
        0.6 & 1.0 &\textcolor{red}{$\bullet $} & \textcolor{red}{$\bullet $} &  \textcolor{grn}{$\bullet $} \\ \hline
        
        0.6 & 0.8 &\textcolor{red}{$\star $} &  \textcolor{grn}{$\bullet $} &  \textcolor{grn}{$\bullet $} \\ \hline
        
        0.6 & 0.6 & \textcolor{black}{$\bullet $} & \textcolor{red}{$\bullet $} &  \textcolor{grn}{$\bullet $} \\ \hline
        
        0.6 & 0.4 & \textcolor{black}{$\bullet $} & \textcolor{red}{$\bullet $} &  \textcolor{grn}{$\star $} \\ \hline
        
        0.6 & 0.2 & \textcolor{black}{$\star $} & \textcolor{red}{$\bullet $} &  \textcolor{black}{$\star $} \\ \hline \hline
        0.4 & 1.0 &\textcolor{red}{$\bullet $} & \textcolor{red}{$\bullet $} &  \textcolor{grn}{$\bullet $} \\ \hline
        
        0.4 & 0.8 &\textcolor{red}{$\bullet $} & \textcolor{red}{$\bullet $} &  \textcolor{grn}{$\bullet $} \\ \hline
        
        0.4 & 0.6 &\textcolor{red}{$\bullet $} &  \textcolor{grn}{$\bullet $} &  \textcolor{grn}{$\bullet $} \\ \hline
        
        0.4 & 0.4 & \textcolor{black}{$\bullet $} & \textcolor{red}{$\bullet $} &  \textcolor{grn}{$\bullet $} \\ \hline
        
        0.4 & 0.2 & \textcolor{black}{$\bullet $}  & \textcolor{red}{$\bullet $} &  \textcolor{grn}{$\bullet $} \\ \hline \hline
        0.2 & 1.0 &\textcolor{red}{$\bullet $} & \textcolor{red}{$\bullet $} &  \textcolor{grn}{$\bullet $} \\ \hline
        
        0.2 & 0.8 &\textcolor{red}{$\bullet $} & \textcolor{red}{$\bullet $} &  \textcolor{grn}{$\bullet $} \\ \hline
        
        0.2 & 0.6 &\textcolor{red}{$\bullet $} & \textcolor{red}{$\bullet $} &  \textcolor{grn}{$\bullet $} \\ \hline
        
        0.2 & 0.4 &\textcolor{red}{$\bullet $} &  \textcolor{grn}{$\bullet $} &  \textcolor{grn}{$\bullet $} \\ \hline
        
        0.2 & 0.2 & \textcolor{black}{$\star $} & \textcolor{red}{$\bullet $} &  \textcolor{grn}{$\bullet $} \\ \hline
        
    \end{tabular}
     \caption{\textbf{\hspace{1cm} For $\varphi_{1} = 50^{\circ}$ and $\varphi_{2} = 300^{\circ}$}}
    \label{alphebet_symtable_50_300}
    \end{subtable}
    \vspace{5pt}
    \caption{The implications for the low scale values of $\theta_{23}$ , $\Delta \mathrm{m^{2}_{atm}}$ and $\mathrm{m_{\beta \beta}}$ for various values of $\mathrm{\alpha}$ and $\mathrm{\beta}$ for $\lambda$ $= \mathrm{\lambda_{HSMU}}$. The coloured symbols represent the following:
\\ \textcolor{black}{$\bullet $}: Parameter value is inside 3-$\sigma$ range \\
\textcolor{red}{$\bullet $}: Parameter value is below its lower bound \\
\textcolor{grn}{$\bullet $}: Parameter value is above its upper bound \\ 
\textcolor{black}{$\star $}: Parameter value is inside, but near the 3-$\sigma$ boundary \\ 
\textcolor{red}{$\star $}: Parameter value is \textit{just} below its lower bound \\ 
\textcolor{grn}{$\star $}: Parameter value is \textit{just} above its lower bound}
    \label{alphebet_symtable_1}
\end{table} \noindent

\subsubsection{Case-1: $\mathrm{\varphi}_1$ $= 50^{\circ}$ and $\mathrm{\varphi}_2$ $= 0^{\circ}$} 

From the case when $\mathrm{\varphi}_1$, $\mathrm{\varphi}_2$ are $50^{\circ}$ and $0^{\circ}$ respectively, Table \ref{alphebet_symtable_50_0}, we can see that red and green symbols occur together for every $\mathrm{\alpha}$, $\mathrm{\beta}$ pairs. Thus we can choose to ignore them for $\lambda$ variations as it signifies that there exists at least one parameter among $\theta_{23}$, $\Delta \mathrm{m^{2}_{atm}}$ and $\mathrm{m_{\beta \beta}}$, which isn't inside its valid 3-$\sigma$ range. Since this happens for every $\mathrm{\alpha}$, $\mathrm{\beta}$ pair, it leaves us with no plausible sets of $\mathrm{\alpha}$, $\mathrm{\beta}$ pairs for $\lambda$ variations. Thus we can rule out the Majorana phases pair ($50^{\circ}, 0^{\circ}$) as a candidate to bring in all six low scale parameters.

\subsubsection{Case-2: $\mathrm{\varphi}_1$ $= 50^{\circ}$ and $\mathrm{\varphi}_2$ $= 300^{\circ}$} 

For the case when $\mathrm{\varphi}_1$, $\mathrm{\varphi}_2$ are $50^{\circ}$ and $300^{\circ}$ respectively, from Tab.~\ref{alphebet_symtable_50_300}, we can see that every time red and green symbols occur together for the same $\mathrm{\alpha}$, $\mathrm{\beta}$ pairs we can choose to ignore them for $\lambda$ variations as it signifies that there exists at least one parameter among $\theta_{23}$, $\Delta \mathrm{m^{2}_{atm}}$ and $\mathrm{m_{\beta \beta}}$, which does not come inside its valid 3-$\sigma$ range. This leaves us with only five plausible sets of $\mathrm{\alpha}$, $\mathrm{\beta}$ pairs. They are as follows: 
\begin{table}[!ht]
    \centering
    \begin{tabular}{|c||c|c|c|c|c|}
     \hline
     \hspace{0.1cm}$\alpha$ \hspace{0.1cm}&\hspace{0.1cm}  1.0 \hspace{0.1cm}&\hspace{0.1cm} 1.0 \hspace{0.1cm}&\hspace{0.1cm} 0.8 \hspace{0.1cm}&\hspace{0.1cm} 0.8 \hspace{0.1cm}&\hspace{0.1cm} 0.6 \hspace{0.1cm} \\ \hline
     $\beta$ &  0.8 & 0.6 & 1.0 & 0.4 & 0.2 \\ \hline
    \end{tabular}
\end{table}

In all these pairs, $\Delta \mathrm{m^{2}_{atm}}$ can be seen below its lower bound. In this set of $\mathrm{\alpha}$-$\mathrm{\beta}$ pairs some of them have low values $\mathrm{\alpha}$ ($\mathrm{\alpha}$ = 04, 0.2). As  discussed earlier, we ignore lower values of $\mathrm{\alpha}$ because $\Delta \mathrm{m^{2}_{atm}}$ doesn't show monotonic increasing or monotonic decreasing trend for variations in $\mathrm{\lambda}$. In the remaining $\mathrm{\alpha}$-$\mathrm{\beta}$ pairs, we will have to decrease $\lambda$ in order to correct $\Delta \mathrm{m^{2}_{atm}}$. But for all of these pairs of $\mathrm{\alpha}$-$\mathrm{\beta}$, $\theta_{23}$ is given by `$\star $'; it represents that the values of $\theta_{23}$ are about to cross the upper bound of 3-$\sigma$ range. Thus decreasing $\lambda$ would increase the values of $\theta_{23}$ and it’ll be out of its valid range. Moreover, if we increase the value of $\lambda$ it’ll further decrease $\Delta \mathrm{m^{2}_{atm}}$ which is already below its valid range. So, keeping both $\Delta \mathrm{m^{2}_{atm}}$ and $\theta_{23}$ inside their valid ranges is not possible for any variation in $\mathrm{\lambda}$. Thus we can, yet again, rule out the Majorana phases pair (300, 50) as a candidate to bring all six low scale parameters inside their respective 3-$\sigma$ ranges.

\subsubsection{Case-3: $\mathrm{\varphi}_1$ $= 200^{\circ}$ and $\mathrm{\varphi}_2$ $= 300^{\circ}$}


From the case when $\mathrm{\varphi}_1$, $\mathrm{\varphi}_2$ are $200^{\circ}$ and $300^{\circ}$ respectively, we have the Tab.~\ref{alphebet_symtable_200_300}. We again choose to ignore simultaneous occurrences of red and green together for $\lambda$ variations. This leaves us with only three plausible sets of $\mathrm{\alpha}$, $\mathrm{\beta}$ pairs. They are as follows
\begin{table}[!ht]
    \centering
    \begin{tabular}{|c||c|c|c|}
     \hline
     \hspace{0.1cm}$\alpha$ \hspace{0.1cm}&\hspace{0.1cm}  1.0 \hspace{0.1cm}&\hspace{0.1cm} 1.0 \hspace{0.1cm}&\hspace{0.1cm} 0.8 \hspace{0.1cm} \\ \hline
     $\beta$ &  0.4 & 0.2 & 0.4 \\ \hline
    \end{tabular}
\end{table}

\begin{table}[!ht]
    \centering
        \begin{subtable}{.5\textwidth} \centering
        \begin{tabular}{|c|c|c|c|c|} 
        \hline
        \hspace{0.1cm}$\mathrm{\alpha}$ \hspace{0.1cm}&\hspace{0.1cm} $\mathrm{\beta}$ \hspace{0.1cm}&\hspace{0.1cm} $\theta_{23}$ \hspace{0.1cm}&\hspace{0.1cm} $\Delta \mathrm{m^{2}_{atm}}$ \hspace{0.1cm}&\hspace{0.1cm} $\mathrm{m_{\beta \beta}}$ \hspace{0.1cm} \\ \hline
        1.0 & 1.0 & \textcolor{grn}{$\bullet $} &  
       \textcolor{red}{$\bullet $} &  
        \textcolor{black}{$\bullet $} \\ \hline
        
        1.0 & 0.8 & \textcolor{grn}{$\star $} &  
       \textcolor{red}{$\bullet $} &  
        \textcolor{black}{$\bullet $} \\ \hline
        
        1.0 & 0.6 & \textcolor{grn}{$\star $} &  
       \textcolor{red}{$\bullet $} & 
        \textcolor{black}{$\bullet $} \\ \hline
        
        1.0 & 0.4 & \textcolor{black}{$\star $} &  
       \textcolor{red}{$\bullet $} &  
        \textcolor{black}{$\bullet $} \\ \hline
        
        1.0 & 0.2 & \textcolor{black}{$\bullet $} &  
       \textcolor{red}{$\bullet $} &  
        \textcolor{black}{$\bullet $} \\ \hline \hline
        0.8 & 1.0 & \textcolor{grn}{$\bullet $} &  
       \textcolor{red}{$\bullet $} &  
        \textcolor{black}{$\bullet $} \\ \hline
        
        0.8 & 0.8 & \textcolor{grn}{$\bullet $} &  
       \textcolor{red}{$\bullet $} &  
        \textcolor{black}{$\bullet $} \\ \hline
        
        0.8 & 0.6 & \textcolor{grn}{$\star $} &  
       \textcolor{red}{$\bullet $} & 
        \textcolor{black}{$\bullet $} \\ \hline
        
        0.8 & 0.4 & \textcolor{black}{$\star $} &  
       \textcolor{red}{$\bullet $} &  
        \textcolor{black}{$\bullet $} \\ \hline
        
        0.8 & 0.2 & \textcolor{black}{$\bullet $} &  
       \textcolor{red}{$\bullet $} &  
        \textcolor{grn}{$\star $} \\ \hline \hline
        0.6 & 1.0 & \textcolor{grn}{$\bullet $} &  
       \textcolor{red}{$\bullet $} &  
        \textcolor{black}{$\bullet $} \\ \hline
        
        0.6 & 0.8 & \textcolor{grn}{$\bullet $} &  
       \textcolor{red}{$\bullet $} &  
        \textcolor{black}{$\bullet $} \\ \hline
        
        0.6 & 0.6 & \textcolor{grn}{$\bullet $} &  
       \textcolor{red}{$\bullet $} &  
        \textcolor{black}{$\bullet $} \\ \hline
        
        0.6 & 0.4 & \textcolor{grn}{$\star $} &  
       \textcolor{red}{$\bullet $} &  
        \textcolor{black}{$\bullet $} \\ \hline
        
        0.6 & 0.2 & \textcolor{black}{$\bullet $} &  
       \textcolor{red}{$\bullet $} &  
        \textcolor{grn}{$\bullet $} \\ \hline \hline
        0.4 & 1.0 & \textcolor{grn}{$\bullet $} &  
       \textcolor{red}{$\bullet $} &  
        \textcolor{black}{$\bullet $} \\ \hline
        
        0.4 & 0.8 & \textcolor{grn}{$\bullet $} &  
       \textcolor{red}{$\bullet $} &  
        \textcolor{black}{$\bullet $} \\ \hline
        
        0.4 & 0.6 & \textcolor{grn}{$\bullet $} &  
       \textcolor{red}{$\bullet $} &  
        \textcolor{black}{$\bullet $} \\ \hline
        
        0.4 & 0.4 & \textcolor{grn}{$\star $} &  
       \textcolor{red}{$\bullet $} &  
        \textcolor{black}{$\bullet $} \\ \hline
        
        0.4 & 0.2 & \textcolor{black}{$\star $}  &  
       \textcolor{red}{$\bullet $} &  
        \textcolor{grn}{$\bullet $} \\ \hline \hline
        0.2 & 1.0 &\textcolor{red}{$\bullet $} &  
       \textcolor{red}{$\bullet $} &  
        \textcolor{grn}{$\bullet $} \\ \hline
        
        0.2 & 0.8 &\textcolor{red}{$\bullet $} &  
       \textcolor{red}{$\bullet $} &  
        \textcolor{grn}{$\bullet $} \\ \hline
        
        0.2 & 0.6 & \textcolor{grn}{$\bullet $} &  
       \textcolor{red}{$\bullet $} &  
        \textcolor{black}{$\bullet $} \\ \hline
        
        0.2 & 0.4 & \textcolor{grn}{$\bullet $} &  
       \textcolor{red}{$\bullet $} &  
        \textcolor{black}{$\bullet $} \\ \hline
        
        0.2 & 0.2 & \textcolor{grn}{$\star $} &  
       \textcolor{red}{$\bullet $} &  
        \textcolor{grn}{$\bullet $} \\ \hline
        
    \end{tabular}
    \caption{\textbf{\hspace{1cm}For $\varphi_{1} = 200^{\circ}$ and $\varphi_{2} = 300^{\circ}$}}
    \label{alphebet_symtable_200_300}
    \end{subtable}%
    \begin{subtable}{.5\textwidth} \centering
        \begin{tabular}{|c|c|c|c|c|} 
        \hline
        \hspace{0.1cm} $\mathrm{\alpha}$ \hspace{0.1cm}&\hspace{0.1cm} $\mathrm{\beta}$ \hspace{0.1cm}&\hspace{0.1cm} $\theta_{23}$ \hspace{0.1cm}&\hspace{0.1cm} $\Delta \mathrm{m^{2}_{atm}}$ \hspace{0.1cm}&\hspace{0.1cm} $\mathrm{m_{\beta \beta}}$ \hspace{0.1cm} \\ \hline
        1.0 & 1.0 & \textcolor{black}{$\star $} &  
       \textcolor{red}{$\bullet $} &  
        \textcolor{black}{$\bullet $} \\ \hline
        
        1.0 & 0.8 & \textcolor{grn}{$\star $} &  
       \textcolor{red}{$\bullet $} &  
        \textcolor{black}{$\bullet $} \\ \hline
        
        1.0 & 0.6 & \textcolor{grn}{$\star $} &  
       \textcolor{red}{$\bullet $} & 
        \textcolor{black}{$\bullet $} \\ \hline
        
        1.0 & 0.4 & \textcolor{grn}{$\star $} &  
       \textcolor{red}{$\bullet $} &  
        \textcolor{black}{$\bullet $} \\ \hline
        
        1.0 & 0.2 & \textcolor{grn}{$\star $} &  
       \textcolor{red}{$\bullet $} &  
        \textcolor{black}{$\bullet $} \\ \hline \hline
        0.8 & 1.0 & \textcolor{black}{$\star $} &  
       \textcolor{red}{$\bullet $} &  
        \textcolor{black}{$\bullet $} \\ \hline
        
        0.8 & 0.8 & \textcolor{black}{$\star $} &  
       \textcolor{red}{$\bullet $} &  
        \textcolor{black}{$\bullet $} \\ \hline
        
        0.8 & 0.6 & \textcolor{grn}{$\star $} &  
       \textcolor{red}{$\bullet $} & 
        \textcolor{black}{$\bullet $} \\ \hline
        
        0.8 & 0.4 & \textcolor{grn}{$\star $} &  
       \textcolor{red}{$\bullet $} &  
        \textcolor{black}{$\bullet $} \\ \hline
        
        0.8 & 0.2 & \textcolor{grn}{$\star $} &  
       \textcolor{red}{$\bullet $} &  
        \textcolor{black}{$\bullet $} \\ \hline \hline
        0.6 & 1.0 & \textcolor{black}{$\star $} &  
       \textcolor{red}{$\bullet $} &  
        \textcolor{black}{$\bullet $} \\ \hline
        
        0.6 & 0.8 & \textcolor{black}{$\star $} &  
       \textcolor{red}{$\bullet $} &  
        \textcolor{black}{$\bullet $} \\ \hline
        
        0.6 & 0.6 & \textcolor{black}{$\star $} &  
       \textcolor{red}{$\bullet $} &  
        \textcolor{black}{$\bullet $} \\ \hline
        
        0.6 & 0.4 & \textcolor{black}{$\star $} &  
       \textcolor{red}{$\bullet $} &  
        \textcolor{black}{$\bullet $} \\ \hline
        
        0.6 & 0.2 & \textcolor{grn}{$\star $} &  
       \textcolor{red}{$\bullet $} &  
        \textcolor{black}{$\bullet $} \\ \hline \hline
        0.4 & 1.0 & \textcolor{black}{$\star $} &  
       \textcolor{red}{$\bullet $} &  
        \textcolor{black}{$\bullet $} \\ \hline
        
        0.4 & 0.8 & \textcolor{black}{$\star $} &  
       \textcolor{red}{$\bullet $} &  
        \textcolor{black}{$\bullet $} \\ \hline
        
        0.4 & 0.6 & \textcolor{black}{$\star $} &  
       \textcolor{red}{$\bullet $} &  
        \textcolor{black}{$\bullet $} \\ \hline
        
        0.4 & 0.4 & \textcolor{black}{$\star $} &  
       \textcolor{red}{$\bullet $} &  
        \textcolor{black}{$\bullet $} \\ \hline
        
        0.4 & 0.2 & \textcolor{black}{$\star $}  &  
       \textcolor{red}{$\bullet $} &  
        \textcolor{black}{$\bullet $} \\ \hline \hline
        0.2 & 1.0 & \textcolor{black}{$\star $} &  
       \textcolor{red}{$\bullet $} &  
        \textcolor{black}{$\star $} \\ \hline
        
        0.2 & 0.8 & \textcolor{black}{$\star $} &  
       \textcolor{red}{$\bullet $} &  
        \textcolor{grn}{$\bullet $} \\ \hline
        
        0.2 & 0.6 & \textcolor{black}{$\star $} &  
       \textcolor{red}{$\bullet $} &  
        \textcolor{grn}{$\bullet $} \\ \hline
        
        0.2 & 0.4 & \textcolor{black}{$\star $} &  
       \textcolor{red}{$\bullet $} &  
        \textcolor{grn}{$\bullet $} \\ \hline
        
        0.2 & 0.2 & \textcolor{black}{$\star $} &  
       \textcolor{red}{$\bullet $} &  
        \textcolor{grn}{$\bullet $} \\ \hline
        
    \end{tabular}
    \caption{\textbf{\hspace{1cm}For $\varphi_{1} = 300^{\circ}$ and $\varphi_{2} = 50^{\circ}$}}
    \label{alphebet_symtable_300_50}
    \end{subtable}
    \vspace{5pt}
    \caption{The implications for the low scale values of $\theta_{23}$ , $\Delta \mathrm{m^{2}_{atm}}$ and $\mathrm{m_{\beta \beta}}$ for various values of $\mathrm{\alpha}$ and $\mathrm{\beta}$ for $\lambda$ $= \mathrm{\lambda_{HSMU}}$. The coloured symbols represent the following: \\
\textcolor{black}{$\bullet $}: Parameter value is inside 3-$\sigma$ range \\
\textcolor{red}{$\bullet $}: Parameter value is below its lower bound \\
\textcolor{grn}{$\bullet $}: Parameter value is above its upper bound \\ 
\textcolor{black}{$\star $}: Parameter value is inside, but near the 3-$\sigma$ boundary \\ 
\textcolor{red}{$\star $}: Parameter value is \textit{just} below its lower bound \\ 
\textcolor{grn}{$\star $}: Parameter value is \textit{just} above its lower bound}
    \label{alphebet_symtable_2}
\end{table}

In all these pairs, $\Delta \mathrm{m^{2}_{atm}}$ is below its lower bound and in order to correct it, we will have to decrease $\lambda$ such that $\Delta \mathrm{m^{2}_{atm}}$ increases. But note that for two of these pairs of $\mathrm{\alpha}$ and $\mathrm{\beta}$, $\theta_{23}$ is given by `$\star $', representing that $\theta_{23}$ is on the edge of its 3-$\sigma$ range. Thus decreasing $\lambda$ would shift $\theta_{23}$ immediately out of its valid range and we won't be able to get $\theta_{23}$ inside when $\Delta \mathrm{m^{2}_{atm}}$ comes inside 3-$\sigma$ range. Upon decreasing $\lambda$ for the remaining case when $\alpha = 1.0$ and $\beta = 0.2$, the same argument applies even though $\theta_{23}$ value is inside 3-$\sigma$ range. That is, it shifts $\theta_{23}$ again out of its valid range and $\Delta \mathrm{m^{2}_{atm}}$ still doesn't come inside. This makes it impossible again to bring all six low scale parameters in. Thus we can again rule out the Majorana phases pair ($200^{\circ}, 300^{\circ}$) as a candidate to bring in all six low scale parameters.

\subsubsection{Case-4: $\mathrm{\varphi}_1$ $= 300^{\circ}$ and $\mathrm{\varphi}_2$ $= 50^{\circ}$}

From the case when $\mathrm{\varphi}_1$, $\mathrm{\varphi}_2$ are $300^{\circ}$ and $50^{\circ}$ respectively, shown in Tab.~\ref{alphebet_symtable_300_50}, we again choose to ignore simultaneous occurrences of red and green together for $\lambda$ variations. This time it leaves us with few more possibilities. It gives us total 13 plausible sets of $\mathrm{\alpha}$, $\mathrm{\beta}$ pairs for which we can then look for $\lambda$ variations. They are as follows
\begin{table}[!ht]
 \centering
 \begin{tabular}{|c||c|c|c|c|c|c|c|c|c|c|c|c|c|}
  \hline
   \hspace{0.1cm}$\alpha$ \hspace{0.1cm}&\hspace{0.1cm} 1.0 \hspace{0.1cm}& \hspace{0.1cm} 0.8 \hspace{0.1cm} & \hspace{0.1cm} 0.8 \hspace{0.1cm} & \hspace{0.1cm} 0.6 \hspace{0.1cm} & \hspace{0.1cm} 0.6 \hspace{0.1cm} & \hspace{0.1cm} 0.6 \hspace{0.1cm} & \hspace{0.1cm} 0.6 \hspace{0.1cm} & \hspace{0.1cm} 0.4 \hspace{0.1cm} & \hspace{0.1cm} 0.4 \hspace{0.1cm} & \hspace{0.1cm} 0.4 \hspace{0.1cm} & \hspace{0.1cm} 0.4 \hspace{0.1cm} & \hspace{0.1cm} 0.4 \hspace{0.1cm} & \hspace{0.1cm} 0.2 \hspace{0.1cm} \\ \hline
   \hspace{0.1cm}$\beta$ \hspace{0.1cm}&\hspace{0.1cm} 1.0 \hspace{0.1cm}& \hspace{0.1cm} 1.0 \hspace{0.1cm} & \hspace{0.1cm} 0.8 \hspace{0.1cm} & \hspace{0.1cm} 1.0 \hspace{0.1cm} & \hspace{0.1cm} 0.8 \hspace{0.1cm} & \hspace{0.1cm} 0.6 \hspace{0.1cm} & \hspace{0.1cm} 0.4 \hspace{0.1cm} & \hspace{0.1cm} 1.0 \hspace{0.1cm} & \hspace{0.1cm} 0.8 \hspace{0.1cm} & \hspace{0.1cm} 0.6 \hspace{0.1cm} & \hspace{0.1cm} 0.4 \hspace{0.1cm} & \hspace{0.1cm} 0.2 \hspace{0.1cm} & \hspace{0.1cm} 1.0 \hspace{0.1cm} \\ \hline
 \end{tabular}
\end{table}

In all these pairs, $\Delta \mathrm{m^{2}_{atm}}$ is below its lower bound. In this set of $\mathrm{\alpha}$-$\mathrm{\beta}$ pairs some of them have low values $\mathrm{\alpha}$ ($\mathrm{\alpha}$ = 0.4, 0.2). As  discussed earlier, we had already rejected lower values of $\mathrm{\alpha}$ because $\Delta \mathrm{m^{2}_{atm}}$ doesn't show a monotonic increasing or a monotonic decreasing trend with variations in $\lambda$.  In the remaining $\mathrm{\alpha}$-$\mathrm{\beta}$ pairs, we will have to decrease $\lambda$ in order to correct $\Delta \mathrm{m^{2}_{atm}}$. But for all of these pairs of $\mathrm{\alpha}$-$\mathrm{\beta}$, $\theta_{23}$ is given by `$\star $'; it represents that the values of $\theta_{23}$ are about to cross the upper bound of 3-$\sigma$ range. Thus decreasing $\lambda$ would increase the values of $\theta_{23}$ and they will be out of their valid range. Moreover, if we increase the value of $\lambda$ it will  further decrease $\Delta \mathrm{m^{2}_{atm}}$ which is already below its valid range. So, keeping both $\Delta \mathrm{m^{2}_{atm}}$ and $\theta_{23}$ inside their allowed  ranges is not possible for any variation in $\mathrm{\lambda}$. Thus we can, yet again, rule out the Majorana phases pair ($300^{\circ}$, $50^{\circ}$) as here too one cannot bring all six low scale parameters inside their respective 3-$\sigma$ ranges. Thus, to conclude, the leptonic mixing angles cannot have a Wolfenstein like hierarchical structure at high scale, as the resulting low scale values of the oscillation parameters and $m_{\beta \beta}$ are incompatible with their allowed experimental ranges.

\section{Conclusions}
\label{concl}

The High Scale Mixing Unification ansatz was one of the appealing possibilities to understand the deeper connection between lepton and quark sectors. The ansatz implies that at some high energy scale, usually taken as the GUT scale, the quark and leptonic mixing matrices are unified. The apparent differences between the two mixing matrices at the low scale of the experiments is then attributed to the large change in the leptonic mixing angles due to RG evolution from the high to the low scale.  Such large RG evolution can be achieved in SuperSymmetric models like MSSM with the SUSY breaking scale in the few TeV range. Since the leptonic mixing angles and the neutrino mass square differences (observed in neutrino oscillations) RG evolve in a correlated fashion, this makes  HSMU ansatz quite predictive. One of its early prediction was that $\theta_{13}$ angle should be non-zero but small~\cite{Mohapatra:2003tw}, a fact later on confirmed by the experiments. In addition correlation between low scale values of $\theta_{13}$ and $\theta_{23}$ were also predicted~\cite{Abbas:2013uqh,Abbas:2014ala}. A generalization of the HSMU ansatz we call the Wolfenstein ansatz was also proposed in literature, where the strict requirement  that at high scale the leptonic and quark mixing matrices are exactly same was relaxed. Instead the ansatz requires that the leptonic mixing angles at high scale should also have a hierarchical form, qualitatively similar to the one observed in the quark sector.  

In this work we have thoroughly investigated both the HSMU and Wolfenstein ansatz, looking at the possibility of their compatibility with the current global fit results. We first started with the more stringent HSMU ansatz and investigated the various possibilities for both Dirac and Majorana neutrinos. For Dirac neutrinos we looked at the possibility of no CP violation as well as the possibility of CP violation in the leptonic sector. In both cases we found that for Dirac neutrinos, the current experimental data is completely incompatible with the HSMU hypothesis. We then looked at the Majorana cases, again looking at the possibility of zero and non-zero Majorana phases. For the case of zero Majorana phases, the results are similar to the Dirac case and is completely incompatible with current global fit results for the neutrino mixing angles. The case of non-zero Majorana phases was analysed in details and although we were able to have four of the neutrino oscillation parameters inside their current  3-$\sigma$ range but we found that all the five oscillation parameters and $m_{\beta \beta}$ can never be simultaneously brought inside their allowed values, even after taking into account the SUSY threshold corrections. 
Thus we concluded that although promising in past, the HSMU hypothesis is no longer compatible with the current global fit results. Finally, we analyzed the various possibilities for the Wolfenstein ansatz and again we found that all the six observables (the five neutrino oscillation ones +  $m_{\beta \beta}$) cannot be brought simultaneously inside their currently allowed ranges, for any choice of the free parameters. Thus, we can finally conclude that the possibility of the leptonic mixing matrix having a hierarchical quark like form at a high scale is no longer viable.  


\begin{acknowledgments}
	The work of RS was supported by the SERB, Government of India grant SRG/2020/002303.
\end{acknowledgments}


\appendix
\section*{Appendix}

\section{Input Parameters in Bottom-Up and Top-Down RG Runnings}
\label{rgrunning}


\begin{itemize} \vspace{-10pt}
    \item $\mathrm{M_{Z}}$ scale = $91.1876$ GeV \vspace{-10pt}
    \item GUT scale = $2 \times 10^{16}$ GeV \vspace{-10pt}
    \item tan$\mathrm{\beta}$ = 55 ($\mathrm{\beta}$ is the ratio of expectation values of Higgs doublets in 2HDM) \vspace{-10pt}
    \item SUSY cutoff scale = 2000 GeV \vspace{-10pt}
    \item Values of gauge coupling constants \vspace{-10pt}
        \begin{itemize}
            \item[$\circ$] Higgs coupling = 0.4615 (at $\mathrm{M_{Z}}$ scale) \&  0.7013 (at GUT scale) \vspace{-10pt}
            \item[$\circ$] Weak coupling = 0.6519 (at $\mathrm{M_{Z}}$ scale) \& 0.6904 (at GUT scale) \vspace{-10pt}
            \item[$\circ$] Strong coupling = 1.2198 (at $\mathrm{M_{Z}}$ scale) \& 0.6928 (at GUT scale)  \vspace{-10pt}
        \end{itemize}
    \item Quark mixing parameters at $\mathrm{M_{Z}}$ scale (mixing angles and Yukawa matrix elements)   (Using Tab.~\ref{ParticleDataGroup:2018ovx} and Ref.~\cite{ParticleDataGroup:2018ovx})  \vspace{-10pt}
    \item Lepton Yukawa matrix elements at $\mathrm{M_{Z}}$ scale  (Using Tab.~\ref{tab:global-fit} and Refs.~\cite{ParticleDataGroup:2018ovx, deSalas:2020pgw})  \vspace{-10pt}
    \item Quark and Lepton CP violation phases 
    \vspace{-10pt}
    \item Self Higgs coupling ($\lambda$) =  0.1291 \vspace{-10pt}
    \item Higgs ground state VEV($\nu$) = $246$ GeV 
\end{itemize}
Values where sources are not mentioned are either computed using \cite{Antusch:2005gp} or taken from \cite{ParticleDataGroup:2018ovx}.

\section{Dominant SUSY Threshold Corrections}
\label{app:threshold}

To compute the low scale threshold corrections we will use the following parameters
\begin{itemize}
 \item[] $\Lambda = \text{SUSY breaking scale}$
 \item[] $\mathrm{g_{w,\Lambda}} = \mathrm{g_{cut}} = \text{Weak coupling at SUSY breaking scale}(\Lambda) = 0.6354$
 \item[] $\theta_{12,\Lambda} = \text{Solar mixing angle at} \Lambda = 34.4153^{\circ}$
 \item[] $\varphi_{1,\Lambda}, \varphi_{2,\Lambda} = \text{Majorana phases at } \Lambda = 15.5832 ^{\circ} \text{ \& } 359.164^{\circ}$
 \item[] $\mathrm{m_{1,\Lambda}}, \mathrm{m_{2,\Lambda}}, \mathrm{m_{3,\Lambda}} = \text{Neutrino masses at } \Lambda = 0.245775 \text{ eV}, 0.245973 \text{ eV \& } 0.248218 \text{ eV}$
\end{itemize}

We also define few functions which we need in threshold corrections formulae.
\begin{itemize}
 \item[] $\mathrm{p(x,y) = \frac{x}{y}}$
 \item[] $\mathrm{q(x,y) = 1 - p^{2}}$
 \item[] $\mathrm{t(x,y) = \frac{g^{2}_{cut}}{32 \pi^{2}}\left[\frac{p^{2}(\Lambda,y)-p^{2}(x,y)}{q(\Lambda,y)q(x,y)} + \frac{q^{2}(x,y)-1}{q^{2}(x,y)}ln(p^{2}(x,y)) - \frac{q^{2}(\Lambda,y)-1}{q^{2}(\Lambda,y)}ln(p^{2}(\Lambda,y)) \right]}$
 \item[] $ \mathrm{m_{com} = \frac{1}{3}(m_1^2+m_2^2+m_3^2)\times 10^{-9}}$
\end{itemize}
Finally, with these input values and functions we can calculate corrections in both $\Delta \mathrm{m}^2_{\mathrm{atm}}$ and $\Delta \mathrm{m}^2_{\mathrm{sol}}$ as follows
\begin{eqnarray}
    \label{solcorr}
    \mathrm{(\Delta m^{2}_{sol})_{corr}} & = & \mathrm{4m^{2}_{com}} \bigg[[\sin^{2}(\theta_{12,\Lambda})\cos(2\varphi_{2,\Lambda})-\cos^{2}(\theta_{12,\Lambda})\cos(2\varphi_{1,\Lambda})] \times t(m_{selectron},m_{Wino}) \nonumber \\ 
& + & [\cos^{2}(\theta_{12,\Lambda})\cos(2\varphi_{2,\Lambda})-\sin^{2}(\theta_{12,\Lambda})\cos(2\varphi_{1,\Lambda})] \times t(m_{\Lambda},m_{Wino}) \bigg] \times 10^{18} \\
\mathrm{(\Delta m^{2}_{atm})_{corr}} & = & \mathrm{4m^{2}_{com}} \bigg[- \cos^{2}(\theta_{12,\Lambda})\cos(2\varphi_{1,\Lambda}) \times t(m_{selectron},m_{Wino}) \nonumber \\ 
& + & [1-\sin^{2}(\theta_{12,\Lambda})\cos(2\varphi_{1,\Lambda})] \times t(m_{\Lambda},m_{Wino}) \bigg] \times 10^{18}
 \label{atmcorr}
\end{eqnarray}
The corrected mass squared differences are given by
\begin{eqnarray}
  \Delta m^{2}_{sol} = (\Delta m^{2}_{sol})_{RG}+(\Delta m^{2}_{sol})_{corr} \\
  \Delta m^{2}_{atm} = (\Delta m^{2}_{atm})_{RG}+(\Delta m^{2}_{atm})_{corr}
   \label{correctedm2}
\end{eqnarray}

\bibliographystyle{utphys}
\bibliography{HSMU_draft}

\providecommand{\href}[2]{#2}\begingroup\raggedright\begin{thebibliography}{10}

\bibitem{ATLAS:2012yve}
{\bfseries ATLAS} Collaboration, G.~Aad {\em et~al.}, ``{Observation of a new
  particle in the search for the Standard Model Higgs boson with the ATLAS
  detector at the LHC},''
  \href{http://dx.doi.org/10.1016/j.physletb.2012.08.020}{{\em Phys. Lett. B}
  {\bfseries 716} (2012) 1--29},
  \href{http://arxiv.org/abs/1207.7214}{{\ttfamily arXiv:1207.7214 [hep-ex]}}.

\bibitem{CMS:2012qbp}
{\bfseries CMS} Collaboration, S.~Chatrchyan {\em et~al.}, ``{Observation of a
  New Boson at a Mass of 125 GeV with the CMS Experiment at the LHC},''
  \href{http://dx.doi.org/10.1016/j.physletb.2012.08.021}{{\em Phys. Lett. B}
  {\bfseries 716} (2012) 30--61},
  \href{http://arxiv.org/abs/1207.7235}{{\ttfamily arXiv:1207.7235 [hep-ex]}}.

\bibitem{Super-Kamiokande:1998kpq}
{\bfseries Super-Kamiokande} Collaboration, Y.~Fukuda {\em et~al.}, ``{Evidence
  for oscillation of atmospheric neutrinos},''
  \href{http://dx.doi.org/10.1103/PhysRevLett.81.1562}{{\em Phys. Rev. Lett.}
  {\bfseries 81} (1998) 1562--1567},
  \href{http://arxiv.org/abs/hep-ex/9807003}{{\ttfamily arXiv:hep-ex/9807003}}.

\bibitem{SNO:2002tuh}
{\bfseries SNO} Collaboration, Q.~R. Ahmad {\em et~al.}, ``{Direct evidence for
  neutrino flavor transformation from neutral current interactions in the
  Sudbury Neutrino Observatory},''
  \href{http://dx.doi.org/10.1103/PhysRevLett.89.011301}{{\em Phys. Rev. Lett.}
  {\bfseries 89} (2002) 011301},
  \href{http://arxiv.org/abs/nucl-ex/0204008}{{\ttfamily
  arXiv:nucl-ex/0204008}}.

\bibitem{DayaBay:2012fng}
{\bfseries Daya Bay} Collaboration, F.~P. An {\em et~al.}, ``{Observation of
  electron-antineutrino disappearance at Daya Bay},''
  \href{http://dx.doi.org/10.1103/PhysRevLett.108.171803}{{\em Phys. Rev.
  Lett.} {\bfseries 108} (2012) 171803},
  \href{http://arxiv.org/abs/1203.1669}{{\ttfamily arXiv:1203.1669 [hep-ex]}}.

\bibitem{RENO:2012mkc}
{\bfseries RENO} Collaboration, J.~K. Ahn {\em et~al.}, ``{Observation of
  Reactor Electron Antineutrino Disappearance in the RENO Experiment},''
  \href{http://dx.doi.org/10.1103/PhysRevLett.108.191802}{{\em Phys. Rev.
  Lett.} {\bfseries 108} (2012) 191802},
  \href{http://arxiv.org/abs/1204.0626}{{\ttfamily arXiv:1204.0626 [hep-ex]}}.

\bibitem{DayaBay:2018yms}
{\bfseries Daya Bay} Collaboration, D.~Adey {\em et~al.}, ``{Measurement of the
  Electron Antineutrino Oscillation with 1958 Days of Operation at Daya Bay},''
  \href{http://dx.doi.org/10.1103/PhysRevLett.121.241805}{{\em Phys. Rev.
  Lett.} {\bfseries 121} no.~24, (2018) 241805},
  \href{http://arxiv.org/abs/1809.02261}{{\ttfamily arXiv:1809.02261
  [hep-ex]}}.

\bibitem{Mohapatra:2003tw}
R.~N. Mohapatra, M.~K. Parida, and G.~Rajasekaran, ``{High scale mixing
  unification and large neutrino mixing angles},''
  \href{http://dx.doi.org/10.1103/PhysRevD.69.053007}{{\em Phys. Rev. D}
  {\bfseries 69} (2004) 053007},
  \href{http://arxiv.org/abs/hep-ph/0301234}{{\ttfamily arXiv:hep-ph/0301234}}.

\bibitem{Mohapatra:2005gs}
R.~N. Mohapatra, M.~K. Parida, and G.~Rajasekaran, ``{Threshold effects on
  quasi-degenerate neutrinos with high-scale mixing unification},''
  \href{http://dx.doi.org/10.1103/PhysRevD.71.057301}{{\em Phys. Rev. D}
  {\bfseries 71} (2005) 057301},
  \href{http://arxiv.org/abs/hep-ph/0501275}{{\ttfamily arXiv:hep-ph/0501275}}.

\bibitem{Mohapatra:2005pw}
R.~N. Mohapatra, M.~K. Parida, and G.~Rajasekaran, ``{Radiative magnification
  of neutrino mixings in split supersymmetry},''
  \href{http://dx.doi.org/10.1103/PhysRevD.72.013002}{{\em Phys. Rev. D}
  {\bfseries 72} (2005) 013002},
  \href{http://arxiv.org/abs/hep-ph/0504236}{{\ttfamily arXiv:hep-ph/0504236}}.

\bibitem{Agarwalla:2006dj}
S.~K. Agarwalla, M.~K. Parida, R.~N. Mohapatra, and G.~Rajasekaran, ``{Neutrino
  Mixings and Leptonic CP Violation from CKM Matrix and Majorana Phases},''
  \href{http://dx.doi.org/10.1103/PhysRevD.75.033007}{{\em Phys. Rev. D}
  {\bfseries 75} (2007) 033007},
  \href{http://arxiv.org/abs/hep-ph/0611225}{{\ttfamily arXiv:hep-ph/0611225}}.

\bibitem{Abbas:2013uqh}
G.~Abbas, S.~Gupta, G.~Rajasekaran, and R.~Srivastava, ``{High Scale Mixing
  Unification for Dirac Neutrinos},''
  \href{http://dx.doi.org/10.1103/PhysRevD.91.111301}{{\em Phys. Rev. D}
  {\bfseries 91} no.~11, (2015) 111301},
  \href{http://arxiv.org/abs/1312.7384}{{\ttfamily arXiv:1312.7384 [hep-ph]}}.

\bibitem{Abbas:2014ala}
G.~Abbas, S.~Gupta, G.~Rajasekaran, and R.~Srivastava, ``{Predictions from High
  Scale Mixing Unification Hypothesis},''
  \href{http://dx.doi.org/10.1103/PhysRevD.89.093009}{{\em Phys. Rev. D}
  {\bfseries 89} no.~9, (2014) 093009},
  \href{http://arxiv.org/abs/1401.3399}{{\ttfamily arXiv:1401.3399 [hep-ph]}}.

\bibitem{Srivastava:2015tza}
R.~Srivastava, ``{Predictions From High Scale Mixing Unification Hypothesis},''
  \href{http://dx.doi.org/10.1007/s12043-015-1163-9}{{\em Pramana} {\bfseries
  86} no.~2, (2016) 425--436},
  \href{http://arxiv.org/abs/1503.07964}{{\ttfamily arXiv:1503.07964
  [hep-ph]}}.

\bibitem{Srivastava:2016fhg}
R.~Srivastava, ``{High Scale Unification of CKM and PMNS Mixing Matrices},''
  \href{http://dx.doi.org/10.1007/978-3-319-25619-1_56}{{\em Springer Proc.
  Phys.} {\bfseries 174} (2016) 369--375}.

\bibitem{Abbas:2015vba}
G.~Abbas, M.~Z. Abyaneh, A.~Biswas, S.~Gupta, M.~Patra, G.~Rajasekaran, and
  R.~Srivastava, ``{High scale mixing relations as a natural explanation for
  large neutrino mixing},''
  \href{http://dx.doi.org/10.1142/S0217751X16500950}{{\em Int. J. Mod. Phys. A}
  {\bfseries 31} no.~17, (2016) 1650095},
  \href{http://arxiv.org/abs/1506.02603}{{\ttfamily arXiv:1506.02603
  [hep-ph]}}.

\bibitem{Abbas:2016qbl}
G.~Abbas, M.~Z. Abyaneh, and R.~Srivastava, ``{Precise predictions for Dirac
  neutrino mixing},'' \href{http://dx.doi.org/10.1103/PhysRevD.95.075005}{{\em
  Phys. Rev. D} {\bfseries 95} no.~7, (2017) 075005},
  \href{http://arxiv.org/abs/1609.03886}{{\ttfamily arXiv:1609.03886
  [hep-ph]}}.

\bibitem{AbdusSalam:2019hov}
S.~S. AbdusSalam, M.~Z. Abyaneh, F.~Ghelichkhani, and M.~Noormandipour,
  ``{Majorana phases in high-scale mixing unification hypotheses},''
  \href{http://dx.doi.org/10.1142/S0217751X21500779}{{\em Int. J. Mod. Phys. A}
  {\bfseries 36} no.~13, (2021) 2150077},
  \href{http://arxiv.org/abs/1912.13508}{{\ttfamily arXiv:1912.13508
  [hep-ph]}}.

\bibitem{Rajasekaran:2019uvd}
G.~Rajasekaran, ``{Does the Wolfenstein form work for the leptonic mixing
  matrix?},'' \href{http://arxiv.org/abs/1907.08380}{{\ttfamily
  arXiv:1907.08380 [hep-ph]}}.

\bibitem{ParticleDataGroup:2018ovx}
{\bfseries Particle Data Group} Collaboration, M.~Tanabashi {\em et~al.},
  ``{Review of Particle Physics},''
  \href{http://dx.doi.org/10.1103/PhysRevD.98.030001}{{\em Phys. Rev. D}
  {\bfseries 98} no.~3, (2018) 030001}.

\bibitem{Casas:1999tg}
J.~A. Casas, J.~R. Espinosa, A.~Ibarra, and I.~Navarro, ``{General RG equations
  for physical neutrino parameters and their phenomenological implications},''
  \href{http://dx.doi.org/10.1016/S0550-3213(99)00781-6}{{\em Nucl. Phys. B}
  {\bfseries 573} (2000) 652--684},
  \href{http://arxiv.org/abs/hep-ph/9910420}{{\ttfamily arXiv:hep-ph/9910420}}.

\bibitem{Antusch:2001ck}
S.~Antusch, M.~Drees, J.~Kersten, M.~Lindner, and M.~Ratz, ``{Neutrino mass
  operator renormalization revisited},''
  \href{http://dx.doi.org/10.1016/S0370-2693(01)01127-3}{{\em Phys. Lett. B}
  {\bfseries 519} (2001) 238--242},
  \href{http://arxiv.org/abs/hep-ph/0108005}{{\ttfamily arXiv:hep-ph/0108005}}.

\bibitem{Antusch:2001vn}
S.~Antusch, M.~Drees, J.~Kersten, M.~Lindner, and M.~Ratz, ``{Neutrino mass
  operator renormalization in two Higgs doublet models and the MSSM},''
  \href{http://dx.doi.org/10.1016/S0370-2693(01)01414-9}{{\em Phys. Lett. B}
  {\bfseries 525} (2002) 130--134},
  \href{http://arxiv.org/abs/hep-ph/0110366}{{\ttfamily arXiv:hep-ph/0110366}}.

\bibitem{Antusch:2002rr}
S.~Antusch, J.~Kersten, M.~Lindner, and M.~Ratz, ``{Neutrino mass matrix
  running for nondegenerate seesaw scales},''
  \href{http://dx.doi.org/10.1016/S0370-2693(02)01960-3}{{\em Phys. Lett. B}
  {\bfseries 538} (2002) 87--95},
  \href{http://arxiv.org/abs/hep-ph/0203233}{{\ttfamily arXiv:hep-ph/0203233}}.

\bibitem{Antusch:2003kp}
S.~Antusch, J.~Kersten, M.~Lindner, and M.~Ratz, ``{Running neutrino masses,
  mixings and CP phases: Analytical results and phenomenological
  consequences},''
  \href{http://dx.doi.org/10.1016/j.nuclphysb.2003.09.050}{{\em Nucl. Phys. B}
  {\bfseries 674} (2003) 401--433},
  \href{http://arxiv.org/abs/hep-ph/0305273}{{\ttfamily arXiv:hep-ph/0305273}}.

\bibitem{Antusch:2005gp}
S.~Antusch, J.~Kersten, M.~Lindner, M.~Ratz, and M.~A. Schmidt, ``{Running
  neutrino mass parameters in see-saw scenarios},''
  \href{http://dx.doi.org/10.1088/1126-6708/2005/03/024}{{\em JHEP} {\bfseries
  03} (2005) 024}, \href{http://arxiv.org/abs/hep-ph/0501272}{{\ttfamily
  arXiv:hep-ph/0501272}}.

\bibitem{Lindner:2005as}
M.~Lindner, M.~Ratz, and M.~A. Schmidt, ``{Renormalization group evolution of
  Dirac neutrino masses},''
  \href{http://dx.doi.org/10.1088/1126-6708/2005/09/081}{{\em JHEP} {\bfseries
  09} (2005) 081}, \href{http://arxiv.org/abs/hep-ph/0506280}{{\ttfamily
  arXiv:hep-ph/0506280}}.

\bibitem{deSalas:2020pgw}
P.~F. de~Salas, D.~V. Forero, S.~Gariazzo, P.~Mart\'\i{}nez-Mirav\'e, O.~Mena,
  C.~A. Ternes, M.~T\'ortola, and J.~W.~F. Valle, ``{2020 global reassessment
  of the neutrino oscillation picture},''
  \href{http://dx.doi.org/10.1007/JHEP02(2021)071}{{\em JHEP} {\bfseries 02}
  (2021) 071}, \href{http://arxiv.org/abs/2006.11237}{{\ttfamily
  arXiv:2006.11237 [hep-ph]}}.

\bibitem{Ma:2014qra}
E.~Ma and R.~Srivastava, ``{Dirac or inverse seesaw neutrino masses with $B-L$
  gauge symmetry and $S_3$ flavor symmetry},''
  \href{http://dx.doi.org/10.1016/j.physletb.2014.12.049}{{\em Phys. Lett. B}
  {\bfseries 741} (2015) 217--222},
  \href{http://arxiv.org/abs/1411.5042}{{\ttfamily arXiv:1411.5042 [hep-ph]}}.

\bibitem{Ma:2015mjd}
E.~Ma, N.~Pollard, R.~Srivastava, and M.~Zakeri, ``{Gauge $B-L$ Model with
  Residual $Z_3$ Symmetry},''
  \href{http://dx.doi.org/10.1016/j.physletb.2015.09.010}{{\em Phys. Lett. B}
  {\bfseries 750} (2015) 135--138},
  \href{http://arxiv.org/abs/1507.03943}{{\ttfamily arXiv:1507.03943
  [hep-ph]}}.

\bibitem{Ma:2015raa}
E.~Ma and R.~Srivastava, ``{Dirac or inverse seesaw neutrino masses from gauged
  $B–L$ symmetry},'' \href{http://dx.doi.org/10.1142/S0217732315300207}{{\em
  Mod. Phys. Lett. A} {\bfseries 30} no.~26, (2015) 1530020},
  \href{http://arxiv.org/abs/1504.00111}{{\ttfamily arXiv:1504.00111
  [hep-ph]}}.

\bibitem{CentellesChulia:2016rms}
S.~Centelles~Chuli\'a, E.~Ma, R.~Srivastava, and J.~W.~F. Valle, ``{Dirac
  Neutrinos and Dark Matter Stability from Lepton Quarticity},''
  \href{http://dx.doi.org/10.1016/j.physletb.2017.01.070}{{\em Phys. Lett. B}
  {\bfseries 767} (2017) 209--213},
  \href{http://arxiv.org/abs/1606.04543}{{\ttfamily arXiv:1606.04543
  [hep-ph]}}.

\bibitem{CentellesChulia:2017koy}
S.~Centelles~Chuli\'a, R.~Srivastava, and J.~W.~F. Valle, ``{Generalized
  Bottom-Tau unification, neutrino oscillations and dark matter: predictions
  from a lepton quarticity flavor approach},''
  \href{http://dx.doi.org/10.1016/j.physletb.2017.07.065}{{\em Phys. Lett. B}
  {\bfseries 773} (2017) 26--33},
  \href{http://arxiv.org/abs/1706.00210}{{\ttfamily arXiv:1706.00210
  [hep-ph]}}.

\bibitem{CentellesChulia:2018gwr}
S.~Centelles~Chuli\'a, R.~Srivastava, and J.~W.~F. Valle, ``{Seesaw roadmap to
  neutrino mass and dark matter},''
  \href{http://dx.doi.org/10.1016/j.physletb.2018.03.046}{{\em Phys. Lett. B}
  {\bfseries 781} (2018) 122--128},
  \href{http://arxiv.org/abs/1802.05722}{{\ttfamily arXiv:1802.05722
  [hep-ph]}}.

\bibitem{CentellesChulia:2018bkz}
S.~Centelles~Chuli\'a, R.~Srivastava, and J.~W.~F. Valle, ``{Seesaw Dirac
  neutrino mass through dimension-six operators},''
  \href{http://dx.doi.org/10.1103/PhysRevD.98.035009}{{\em Phys. Rev. D}
  {\bfseries 98} no.~3, (2018) 035009},
  \href{http://arxiv.org/abs/1804.03181}{{\ttfamily arXiv:1804.03181
  [hep-ph]}}.

\bibitem{Bonilla:2018ynb}
C.~Bonilla, S.~Centelles-Chuli\'a, R.~Cepedello, E.~Peinado, and R.~Srivastava,
  ``{Dark matter stability and Dirac neutrinos using only Standard Model
  symmetries},'' \href{http://dx.doi.org/10.1103/PhysRevD.101.033011}{{\em
  Phys. Rev. D} {\bfseries 101} no.~3, (2020) 033011},
  \href{http://arxiv.org/abs/1812.01599}{{\ttfamily arXiv:1812.01599
  [hep-ph]}}.

\bibitem{CentellesChulia:2019xky}
S.~Centelles~Chuli\'a, R.~Cepedello, E.~Peinado, and R.~Srivastava,
  ``{Systematic classification of two loop $d$ = 4 Dirac neutrino mass models
  and the Diracness-dark matter stability connection},''
  \href{http://dx.doi.org/10.1007/JHEP10(2019)093}{{\em JHEP} {\bfseries 10}
  (2019) 093}, \href{http://arxiv.org/abs/1907.08630}{{\ttfamily
  arXiv:1907.08630 [hep-ph]}}.

\bibitem{CentellesChulia:2020dfh}
S.~Centelles~Chuli\'a, R.~Srivastava, and A.~Vicente, ``{The inverse seesaw
  family: Dirac and Majorana},''
  \href{http://dx.doi.org/10.1007/JHEP03(2021)248}{{\em JHEP} {\bfseries 03}
  (2021) 248}, \href{http://arxiv.org/abs/2011.06609}{{\ttfamily
  arXiv:2011.06609 [hep-ph]}}.

\bibitem{Chulia:2021jgv}
S.~C. Chuli\'a, ``{Theory and phenomenology of Dirac neutrinos},''
  \href{http://arxiv.org/abs/2110.15755}{{\ttfamily arXiv:2110.15755
  [hep-ph]}}.

\bibitem{Planck:2018vyg}
{\bfseries Planck} Collaboration, N.~Aghanim {\em et~al.}, ``{Planck 2018
  results. VI. Cosmological parameters},''
  \href{http://dx.doi.org/10.1051/0004-6361/201833910}{{\em Astron. Astrophys.}
  {\bfseries 641} (2020) A6}, \href{http://arxiv.org/abs/1807.06209}{{\ttfamily
  arXiv:1807.06209 [astro-ph.CO]}}. [Erratum: Astron.Astrophys. 652, C4
  (2021)].

\bibitem{KATRIN:2021uub}
{\bfseries KATRIN} Collaboration, M.~Aker {\em et~al.}, ``{Direct neutrino-mass
  measurement with sub-electronvolt sensitivity},''
  \href{http://dx.doi.org/10.1038/s41567-021-01463-1}{{\em Nature Phys.}
  {\bfseries 18} no.~2, (2022) 160--166},
  \href{http://arxiv.org/abs/2105.08533}{{\ttfamily arXiv:2105.08533
  [hep-ex]}}.

\bibitem{Weinberg:1979sa}
S.~Weinberg, ``{Baryon and Lepton Nonconserving Processes},''
  \href{http://dx.doi.org/10.1103/PhysRevLett.43.1566}{{\em Phys. Rev. Lett.}
  {\bfseries 43} (1979) 1566--1570}.

\bibitem{KamLANDZen}
{\bfseries KamLAND-Zen} Collaboration, A.~Gando {\em et~al.}, ``{Search for
  Majorana Neutrinos near the Inverted Mass Hierarchy Region with
  KamLAND-Zen},'' \href{http://dx.doi.org/10.1103/PhysRevLett.117.082503}{{\em
  Phys. Rev. Lett.} {\bfseries 117} no.~8, (2016) 082503},
  \href{http://arxiv.org/abs/1605.02889}{{\ttfamily arXiv:1605.02889
  [hep-ex]}}. [Addendum: Phys.Rev.Lett. 117, 109903 (2016)].

\end{thebibliography}\endgroup
\end{document}